\documentclass[sigconf, nonacm, pdfa]{acmart}
\usepackage[a-2b,mathxmp]{pdfx}
\newcommand\vldbdoi{XX.XX/XXX.XX}
\newcommand\vldbpages{XXX-XXX}
\newcommand\vldbvolume{18}
\newcommand\vldbissue{8}
\newcommand\vldbyear{2025}
\newcommand\vldbauthors{\authors}
\newcommand\vldbtitle{\shorttitle} 
\newcommand\vldbavailabilityurl{https://github.com/shoupzwu/GRASP}
\newcommand\vldbpagestyle{empty} 

\usepackage{colorprofiles}
\hypersetup{pdfstartview=}

\usepackage{bm}
\usepackage{url}
\usepackage{xcolor}
\usepackage{array}   
\usepackage{graphicx}
\usepackage{makecell}
\usepackage[noend]{algorithmic}
\usepackage{caption}
\usepackage{hyperref}

\usepackage{diagbox}
\usepackage{multirow}
\usepackage{enumitem}
\usepackage{subcaption}
\usepackage{tabularx, environ}
\usepackage{color}
\usepackage{longtable}
\usepackage{supertabular}

\usepackage{mdframed}
\newmdenv[linecolor=black, backgroundcolor=gray!10]{boxedalgorithm}
\usepackage[ruled,vlined]{algorithm2e}
\usepackage{fancyvrb}
\usepackage{xspace}
\usepackage{color} 
\usepackage[utf8]{inputenc}
\usepackage{balance} 
\usepackage{eqparbox}
\usepackage{etoolbox} 
\usepackage{pgfplots}
\usepackage{balance} 
\usepackage{subcaption} 
\usepackage{marginnote}
\usepackage{caption}
\usepackage{listings}
\usepackage{amsmath}
\usepackage{xspace}
\usepackage{fancyhdr}
\usepackage{mathtools}
\usepackage{amsthm}
\usepackage{amssymb} 
\usepackage{pifont}
\usepackage{caption}
\usepackage{amsthm}
\usepackage{enumitem}
\usepackage{tikz}
\usepackage{textcomp}
\usetikzlibrary{arrows.meta, positioning, shapes.geometric}
\usetikzlibrary{positioning, fit}
\usetikzlibrary{matrix, decorations.pathreplacing,calligraphy}

\definecolor{darkblue}{rgb}{0.0, 0.0, 0.55}
\definecolor{maroon}{rgb}{0.5, 0.0, 0.0}
\usepackage[normalem]{ulem}

\newtheorem{exmp}{Example}[section]
\newcommand{\eat}[1]{}

\definecolor{keycolor}{rgb}{0.0, 0.0, 0.5}   
\definecolor{strcolor}{rgb}{0.1, 0.5, 0.1}   
\definecolor{condcolor}{rgb}{0.9, 0.4, 0.0} 

\newtheoremstyle{nodefindent}
  {3pt}
  {3pt}
  {\normalfont}
  {0pt}
  {\bfseries}
  {.}
  {.5em}
  {}

\theoremstyle{nodefindent}

\newtheorem{definition}{Definition}[section]

\newcounter{nalg}[section] 
\renewcommand{\thenalg}{\arabic{nalg}} 
\DeclareCaptionLabelFormat{algocaption}{Algorithm \thenalg} 

\lstnewenvironment{algo}[1][] 
{   
    \refstepcounter{nalg} 
    \captionsetup{labelformat=algocaption,labelsep=colon} 
    \lstset{ 
        mathescape=true,
        frame=tB,
        numbers=left, 
        numberstyle=\tiny,
        basicstyle=\scriptsize, 
        keywordstyle=\color{black}\bfseries\em,
        keywords={,input, output, return, datatype, function, in, if, else, foreach, while, begin, end, } 
        numbers=left,
        xleftmargin=.04\textwidth,
        #1 
    }
}
{}

\tikzset{%
  every neuron/.style={
    circle,
    draw,
    minimum size=1cm
  },
  neuron missing/.style={
    draw=none, 
    scale=4,
    text height=0.333cm,
    execute at begin node=\color{black}$\vdots$
  },
}

\definecolor{mycolor}{RGB}{255,0,0} 
\definecolor{palebrown}{rgb}{0.6, 0.46, 0.33}

\newcommand{\neucdf}{\mbox{$\textsc{NeuroCDF}$}\xspace}
\newcommand{\arcdf}{\mbox{$\textsc{ArCDF}$}\xspace}
\newcommand{\name}{\mbox{$\textsc{GRASP}$}\xspace}

\newcommand{\mscnwd}{\mbox{$\text{MSCN}_{\text{w/Data}}$}\xspace}
\newcommand{\mscnnod}{\mbox{$\text{MSCN}_{\text{w/oData}}$}\xspace}

\numberwithin{assumption}{section}

\definecolor{darkgreen}{rgb}{0.0, 0.5, 0.0} 

\begin{document}

\title{Data-Agnostic Cardinality Learning from Imperfect Workloads}

\author{Peizhi Wu}
\authornote{Most of the work was done during Peizhi Wu's internship at ByteDance.}
\affiliation{%
  \institution{University of Pennsylvania}
}
\email{pagewu@cis.upenn.edu}

\author{Rong Kang}
\affiliation{%
  \institution{ByteDance}
}
\email{kangrong.cn@bytedance.com}

\author{Tieying Zhang}
\authornote{Corresponding author.} 
\affiliation{%
  \institution{ByteDance}
}
\email{tieying.zhang@bytedance.com}

\author{Jianjun Chen}
\affiliation{%
  \institution{ByteDance}
}
\email{jianjun.chen@bytedance.com}

\author{Ryan Marcus}
\affiliation{%
  \institution{University of Pennsylvania}
}
\email{rcmarcus@cis.upenn.edu} 

\author{Zachary G. Ives}
\affiliation{%
  \institution{University of Pennsylvania}
}
\email{zives@cis.upenn.edu}

\begin{abstract}
Cardinality estimation (CardEst) is a critical aspect of query optimization. Traditionally, it leverages statistics built directly over the data. However, organizational policies (e.g., regulatory compliance) may restrict global data access. Fortunately, \emph{query-driven} cardinality estimation can learn CardEst models using query workloads. However, existing query-driven models often require access to data or summaries for best performance, and they assume \emph{perfect} training workloads with complete and balanced join templates (or join graphs). Such assumptions rarely hold in real-world scenarios, in which join templates are incomplete and imbalanced.

We present GRASP, a \emph{data-agnostic} cardinality learning system designed to work under these real-world constraints. GRASP's \emph{compositional design} generalizes to unseen join templates and is robust to join template imbalance. It also introduces a new per-table CardEst model that handles value distribution shifts for range predicates, and a novel \emph{learned count sketch} model that captures join correlations across base relations. Across three database instances, we demonstrate that GRASP consistently outperforms existing query-driven models on imperfect workloads, both in terms of estimation accuracy and query latency. Remarkably, GRASP achieves performance comparable to, or even surpassing, traditional approaches built over the underlying data on the complex CEB-IMDb-full benchmark --- despite operating without any data access and using only 10\% of all possible join templates.
\end{abstract}

\maketitle

\pagestyle{\vldbpagestyle}

\begingroup\small\noindent\raggedright\textbf{PVLDB Reference Format:}\\
\vldbauthors. \vldbtitle. PVLDB, \vldbvolume(\vldbissue): \vldbpages, \vldbyear.\\
\href{https://doi.org/\vldbdoi}{doi:\vldbdoi}
\endgroup
\vspace{-0.3em}
\begingroup
\renewcommand\thefootnote{}\footnote{\noindent
This work is licensed under the Creative Commons BY-NC-ND 4.0 International License. Visit \url{https://creativecommons.org/licenses/by-nc-nd/4.0/} to view a copy of this license. For any use beyond those covered by this license, obtain permission by emailing \href{mailto:info@vldb.org}{info@vldb.org}. Copyright is held by the owner/author(s). Publication rights licensed to the VLDB Endowment. \\
\raggedright Proceedings of the VLDB Endowment, Vol. \vldbvolume, No. \vldbissue\ %
ISSN 2150-8097. \\
\href{https://doi.org/\vldbdoi}{doi:\vldbdoi} \\
}\addtocounter{footnote}{-1}\endgroup

\ifdefempty{\vldbavailabilityurl}{}{
\vspace{.3cm}
\begingroup\small\noindent\raggedright\textbf{PVLDB Artifact Availability:}\\
The source code, data, and/or other artifacts have been made available at \url{\vldbavailabilityurl}.
\endgroup
}


\section{Introduction\label{section:intro}}

At enterprise scale and beyond, data management and access may be \emph{compartmentalized}, due to (1) organizational structure and implementation (e.g., different groups adopt incompatible database platforms or access control mechanisms), (2) limited-access agreements between groups (e.g., one organization only grants access through restricted query APIs), and/or (3) regulatory policy and the principle of least privilege (e.g., due to HIPAA or FERPA, certain data fields may be protected).  In such settings, query optimization remains essential for performance, despite the fact some data is kept private\footnote{Note that by ``privacy'', we refer to constraints on what data can be seen by other users or systems. The restrictions do not guarantee differential privacy~\cite{dwork2006differential}.}: the organization may need to execute federated queries~\cite{giannakouris2022building}, provide a service that abstracts over alternative DBMS platforms~\cite{gadepally2016bigdawg}, or even develop query optimization-as-a-service~\cite{jindal2022query}.  Consider our motivating use case involving two real-world scenarios at ByteDance. First, within the organization, numerous internal business units require query optimization services to enhance performance. However, these units often handle sensitive data, such as TikTok user profiles and e-commerce transactions, which are subject to strict data privacy constraints. Consequently, sharing this data across departments is restricted. Additionally, ByteDance's cloud-native services provide query services, but many users are hesitant to allow access to their personal data.

Traditional query optimization relies on cardinality estimation (CardEst) that leverages statistics built directly over base relations.  A potentially promising alternative is to instrument \emph{query workloads} and their cardinality results, which typically raise fewer privacy concerns than direct data access, and are often shared more readily by users with database vendors for optimizing query performance. Importantly, pairs of SQL queries and their resultant cardinalities can be collected with minimal impact on system performance. This facilitates the feasibility of developing learned \emph{query-driven} CardEst models based primarily on query workloads, circumventing the need for direct access to underlying data.
Recent studies~\cite{negi2023robust, wu2024practical} also demonstrate that machine learning (ML)-based query-driven approaches, which learn a regression model that predicts the cardinality for an input query, have the potential to consistently outperform traditional histogram-based methods. This is because traditional methods make assumptions like independence and uniformity of data distributions, whereas learning-based techniques can model data correlations and skewness.  These lessons motivate us to adopt query-driven CardEst techniques, which do not access the underlying data while enhancing estimation accuracy.

\noindent {\color{black} \textbf{Limitations of existing approaches.}} However, it is challenging to apply existing ML-based query-driven CardEst methods to our real-world settings {\color{black}for three main reasons}.

First, most of them~\cite{kipf2019learned, dutt2020efficiently} are \emph{not} purely query-driven as they still require data-derived metadata (\textit{e.g.,} samples and histogram estimates) for better generalization performance. 

\begin{table}
  \caption{Problem settings for query-driven cardinality estimation in the literature versus production environments.} \label{tab:setting_compare}
\scalebox{0.93}{
  \begin{tabularx}{0.5\textwidth}{c|c|c|c|c}
    \hline
\thead{Problem} & \thead{Data} & \thead{Coverage of} & \thead{Distribution of} &  \thead{Distribution of}  \\
   \thead{Setting} & \thead{Access} & \thead{Join Templates} & \thead{Join Templates} & \thead{Predicate Values}\\
 \hline
   \thead{in the literature}& \thead{Yes}  & \thead{Complete}  & \thead{Balanced}  & \thead{Static}\\
  \hline
  \thead{in production}& \thead{No}  & \thead{Incomplete}  & \thead{Imbalanced}  &\thead{Shifting}  \\
  \hline
  \end{tabularx}}
\end{table}

  \begin{figure}
\scalebox{0.85}{
  \begin{tikzpicture}[node distance=1cm and 0.5cm]
\tikzset{
    box/.style={rectangle, draw, fill=yellow!20, minimum width=1.5cm, minimum height=0.5cm},
    state/.style={ellipse, draw, minimum width=1.5cm, minimum height=0.5cm},
    bluebox/.style={rectangle, draw, fill=blue!10, minimum width=1.5cm, minimum height=0.5cm},
    arrow/.style={->, thick, black}
}
    \node[state] (A) {$A$};
    \node[state, below=of A,  yshift=0.8cm] (B) {$B$};
    \node[state, below=of B,  yshift=0.8cm] (AB) {$A\Join B$};
    
    \node[box, right=of A] (Abox) {$A$};
    \node[box, right=of B] (Bbox) {$B$};
    \node[box, right=of AB] (ABbox) {$A\Join B$};
    
    \draw[arrow] (A) -- (Abox);
    \draw[arrow] (B) -- (Bbox);
    \draw[arrow] (AB) -- (ABbox);

    \node[below=0.1cm of AB, xshift=1.1cm, font=\small] (desc1) {\textbf{LW-NN: Separate Model for Each}};

    \node[state, right=of Abox] (MSCNA) {$A$};
    \node[state, right=of Bbox] (MSCNB) {$B$};
    \node[state, right=of ABbox] (MSCNAB) {$A\Join B$};

    \node[box, minimum height=1cm, right=of MSCNB, font=\small] (mscnmodel) {One Model For All};

      \draw[arrow] (MSCNA) -- (mscnmodel);
    \draw[arrow] (MSCNB) -- (mscnmodel);
    \draw[arrow] (MSCNAB) -- (mscnmodel);

     \node[below=0.1cm of MSCNAB, xshift=1.6cm, font=\small] (desc1) {\textbf{MSCN: One Model for All}};

    \node[bluebox, below=of AB, ] (blueA) {$A$};
    \node[bluebox, below=of blueA,  yshift=0.4cm] (blueB) {$B$};

   \node[state, below=of ABbox, yshift=0.3cm] (graspa) {$A$};
    \node[state, below=of graspa, yshift=0.7cm] (graspab) {$A\Join B$};
    \node[state, below=of graspab, yshift=0.7cm] (graspb) {$B$};\

    \node[box, right=of blueA, xshift=2cm] (graspceA) {A};
    \node[box, right=of blueB, xshift=2cm] (graspceB) {$B$};

    \draw[arrow] (graspa) -- (graspceA);
    \draw[arrow] (graspb) -- (graspceB);
    
    \draw[arrow] (graspab) -- (graspceA);
    \draw[arrow] (graspab) -- (graspceB);
      \draw[arrow] (graspab) -- (blueA);
    \draw[arrow] (graspab) -- (blueB);

     \node[below=0.1cm of graspb, xshift=0cm, font=\small] (desc1) {\textbf{\name: Compositional Model Design}};
     
    \node[state, right=of graspceA, minimum width=0.7cm, minimum height=0.3cm, yshift=0.1cm] (query-legend){};
     \node[box, below=of query-legend, minimum width=0.7cm, minimum height=0.3cm, yshift=0.7cm] (card-legend){};
      \node[bluebox, below=of card-legend,  minimum width=0.7cm, minimum height=0.3cm, yshift=0.7cm] (jk-legend){};

    \node[right=0.1cm of query-legend, font=\small] (l1) {Join template};
      \node[right=0.1cm of card-legend, font=\small] (l2) {CardEst model};
    \node[right=0.1cm of jk-legend, font=\small] (l3) {LCS model};
\draw[rounded corners] (5.1, -2.65) rectangle (8,-4.7);




    
\end{tikzpicture}
  }
\caption{\mbox{Join handling in existing query-driven methods}} \label{fig:joinhandling}
\end{figure}
 
Second, they assume \emph{perfect training workloads} in which the join templates (\textit{i.e.,} join graphs that connect base tables\footnote{{\color{black}Join templates and join graphs are interchangeable throughout this paper.}}) are complete and balanced. As demonstrated in \S~\ref{section.analysis}, these assumptions may not hold in real-world environments. Specifically, apart from data inaccessibility, as shown in Table~\ref{tab:setting_compare}, production workloads are inherently \emph{imperfect}, characterized by \emph{incomplete and imbalanced join templates}. This issue is especially problematic in business workloads, which may introduce new join templates over time (Figure~\ref{subfig:business_join_shift}). Even if join templates remain constant, for a query from seen join templates, cost models must estimate cardinalities for all possible subqueries --- including those with join templates not seen in the workload\footnote{In real-world workloads, such as those collected at Bytedance, we only observe the ultimate cardinality of each query without access to the cardinalities of its subqueries.} --- to make accurate cost predictions.
Unfortunately, existing query-driven models suffer in these scenarios. {\color{black} Specifically,  for \textbf{join template incompleteness}, as shown in Figure~\ref{fig:joinhandling}, they handle joins by either encoding each join in the query encoding individually~\cite{kipf2019learned} or by training separate models for each join template~\cite{dutt2020efficiently}. However, neither approach generalizes well to unseen join templates. While join bitmaps~\cite{negi2023robust} improve their generalizability to unseen join templates, constructing join bitmaps requires data access. Furthermore, for \textbf{join template imbalance}, the lack of consistency across join templates in existing methods means that knowledge learned from the ``majority'' join templates (with more training queries) cannot be transferred to the ``minority'' join templates (with very few training queries).}

Third, even for the same query templates (with fixed join graph and queried columns), production workloads often experience shifts (Figure~\ref{subfig:business_query_shift}) in value/literal distributions (aka., \emph{value distribution shifts}). {\color{black} However, existing query-driven approaches rely heavily on data information (\textit{e.g.,} samples, histograms) to handle value distribution shifts~\cite{kipf2019learned, negi2023robust}. Nevertheless, without such data information (a key constraint in this paper), these models are prone to overfitting to the training query distribution, resulting in poor performance when tested on queries from different distributions~\cite{negi2023robust,wu2024modeling}.}

\smallskip
\noindent \textbf{Contributions.} Motivated by these challenges, 
this paper makes contributions as follows.

\begin{itemize}  [leftmargin=*]

 \item  We summarize \textbf{real query workloads in production,} leading to a \textbf{new problem setting} for CardEst: \emph{data-agnostic cardinality learning from imperfect workloads} (\S~\ref{section.problem_formulation}).

\item We develop \name, a \emph{truly data-agnostic} CardEst system that does not require data access and provides robust and generalizable estimates over imperfect workloads. \name develops a design innovation (\textbf{D1}) with two components (\textbf{C1}, \textbf{C2}): 

\begin{enumerate}[label=]
    \item[\textbf{D1.}]\textbf{A general and compositional design} that handles incomplete and imbalanced join templates, using the notion of \textbf{compositional generalization} (\S~\ref{section.compo_design}).
     \item[\textbf{C1.}]\textbf{A new query-driven per-table CardEst model} (\arcdf) that is robust to changes in value distributions for range predicates (\S~\ref{section.arcdf}). 
    {\color{black} \arcdf is inspired by \neucdf, which introduces the CDF modeling paradigm for CardEst. However, \neucdf, as a framework, does not specify the CDF prediction model; and existing attempts, as discussed in ~\cite{wu2024practical}, may face issues with negative estimates. \arcdf mitigates these challenges by introducing \emph{a new CDF prediction model} that utilizes a deep autoregressive model and enforces monotonicity through \emph{monotonic piecewise splines}.

    }
    \item[\textbf{C2.}]\textbf{A novel {query-driven learned count sketch (LCS)} model} that captures join correlations across base relations (\S~\ref{section.alv}). {\color{black} Instead of  prior count sketches built over the data, the LCS model learns from queries to output \emph{low-dimensional representations} that effectively approximate the dot products of join key distributions in the results of per-table subqueries}.
\end{enumerate}



\item  
We validate that \name, with no data or statistics, achieves both \textbf{generalizable and robust CardEst accuracy} and \textbf{reduced query latency} over imperfect training workloads. Notably, on the complex CEB-IMDb-full benchmark~\cite{negi2021flow} with up to 16-way joins, \name achieves comparable or even superior performance to traditional methods with access to the data (\S~\ref{section.exps}). 
We use \emph{only} 10\% of all possible join templates.
\end{itemize}

\noindent
\name performs well with perfect workloads. However, since we developed it based on the challenges of real-world scenarios, we focus on evaluating its performance under those conditions to highlight the \emph{practical utility} of \name.

\smallskip
\noindent \textbf{Data updates/shifts.} In production environments, we observe that user data typically remains static throughout a week, and production systems often accumulate a significant number of queries weekly. For example, businesses we can access in ByteDance generates, on average, over 2 million queries per workload (each corresponding to a database instance) each week.
Therefore, we adopt a pragmatic approach to manage data distribution shifts: we retrain \name weekly using the queries collected during that period.

\section{PRELIMINARIES}
This section first introduces notations and concepts, and then summarizes real-world production workloads we collected at ByteDance.

\vspace{-0.5em}
\subsection{Definitions\label{section:definition}}

\begin{definition}[\textit{Database Instance}] A database instance \(\mathcal{DB}\) comprises a set of base relations/tables \(\{T_j\}_{j=1}^m\), where \(m\) denotes the number of tables in \(\mathcal{DB}\). Each table \(T_j\) includes a set of columns or attributes \(\{A_i\}_{i=1}^n\), where \(n\) represents the number of attributes per table. We define the cardinality $|T_j|$ of each table $T_j$ as the number of total tuples in $T_j$. 
We also define the domain of an attribute \(A\), denoted as \(\text{dom}(A)\), as the set of all distinct values in \(A\). 

\end{definition}

{\color{black}
\begin{definition}
[\textit{Query}]  \label{def:query} A query is a structured request to retrieve data from a database based on specified operations and conditions. This paper focuses on SPJ queries with \emph{inner equi-joins}, following most  learned CardEst work~\cite{wu2023factorjoin, kim2024asm, deepdb, yang2020neurocard}. \name is capable of handling chain joins, star joins, and self-joins, all of which we evaluate in the paper. While \name can also be extended to handle cyclic joins under the join key independence assumption (as in~\cite{wu2023factorjoin}), we did not evaluate cyclic joins, as most benchmarks focus on acyclic joins. For supported predicates, we focus on equality ($=$), range ($<, \leq, >, \geq$), string matching (\texttt{LIKE}), containment  (\texttt{IN}), and null-checking  (\texttt{NULL}) predicates. These supported predicates are consistent with most existing learned CardEst work and are included in the benchmarks we evaluate in this paper. 

Other features (\textit{e.g.,} \texttt{Group-By}, \texttt{Distinct}) also impacts query cardinalities. While existing techniques~\cite{kipf2019estimating} could be integrated with \name to address additional features, they fall outside the scope of this work. Moreover, these features are not prevalent in the real-world queries we collected from ByteDance: 82\% of queries we collected fit within the scope of this paper.

\end{definition}
}

\begin{definition}[\textit{Join Template/Join Graph}] \label{def:join_template} A join template, also referred to as a join graph, connects base relations through their join keys. It consists of nodes and edges, where each node represents a base relation and each edge represents a join operation between two relations based on join keys. {\color{black} Here, a {join key} is defined as a pair of attributes from two relations that are used to establish a condition for joining tuples from those relations. This paper considers both foreign key-primary key (FK-PK) and FK-FK equi-joins}. 
\end{definition}


\begin{definition}[\textit{Query Template}] \label{def:query_template} 
A query template is a {\color{black}parameterized} query pattern that maintains consistent {\color{black}or fixed} join template {\color{black}\emph{as well as other predicates outside of the joining conditions}, varying \emph{only} in the values or literals specified within non-join predicates. This paper does not consider nested queries, following most of learned CardEst work~\cite{wu2023factorjoin, kim2024asm, deepdb, yang2020neurocard}.}
\end{definition}

\begin{definition}[\textit{Cardinality Estimation}] \label{definition:cardest}
    Given a database instance $\mathcal{DB}$ and query $q$ over $\mathcal{DB}$, the goal of cardinality estimation is to predict the cardinality $c(q)$, \textit{i.e.,} the number of tuples that satisfy $q$. Another equivalent term, selectivity, is the ratio $c(q)/|T|$, where $|T|$ is the table cardinality or the join size if $q$ is a join query. {\color{black} In query optimization, the optimizer needs to estimate the cardinality for each subquery of $q$, as each subquery can be considered a query.}
\end{definition}
{ \color{black}


\begin{definition}[\textit{Template Coverage Ratio}] \label{def:tcr} 
 Let $\mathcal{T}$ be the set of all possible join templates of a schema, and 
$\mathcal{T}_{\text{train}} \subset \mathcal{T}$ be the set of templates observed during training. We define \textbf{Template Coverage Ratio (TCR)} as $
        \mathrm{TCR} = \frac{\lvert \mathcal{T}_{\text{train}} \rvert}{\lvert \mathcal{T} \rvert}.$
    A lower TCR indicates that more join templates are missing from the training workload.
\end{definition}

\begin{definition}[\textit{Class Imbalance Ratio}]\label{def:cir}   
 Let $n_t$ be the number of training queries for a join template 
$t \in \mathcal{T}_{\text{train}}$. We define  \textbf{Class Imbalance Ratio (CIR):} $ \mathrm{CIR} = \frac{\max_{t \in \mathcal{T}_{\text{train}}} n_t} {\min_{t \in \mathcal{T}_{\text{train}}} n_t}.$
 A higher CIR indicates a greater imbalance between join templates.
\end{definition}

\begin{definition}[\textit{Granularity}]\label{def:granularity} 
We define \textbf{Granularity} as the \emph{range size} for a query $q$ on a specific \emph{numerical} attribute $A_i$ that supports \emph{range predicates/filters}. For example, let $\ell_i$ and $r_i$ be the lower and upper bounds of $q$'s filter over $A_i$. Then: $\mathrm{Granularity}(q) = r_i - \ell_i$. \label{definition.granularity}
\end{definition}

}

\vspace{-1em}
\subsection{\mbox{Analysis of Production Workloads}~\label{section.analysis}}

This section uses real workloads collected from ByteDance, a technology company operating a range of social media platforms, to illustrate the characteristics (Figure~\ref{fig:bytedance_workload}) of RDBMS workloads over 30 days with more than 35,000 applications. {\color{black} In Figure~\ref{subfig:business_imcomplete}~\ref{subfig:business_imbalance}, a workload index corresponds to the query workload of an internal business.} The key features of these workloads are detailed as follows:

\begin{figure}
  \centering
  \begin{subfigure}{0.23\textwidth}
    \includegraphics[width=\textwidth]{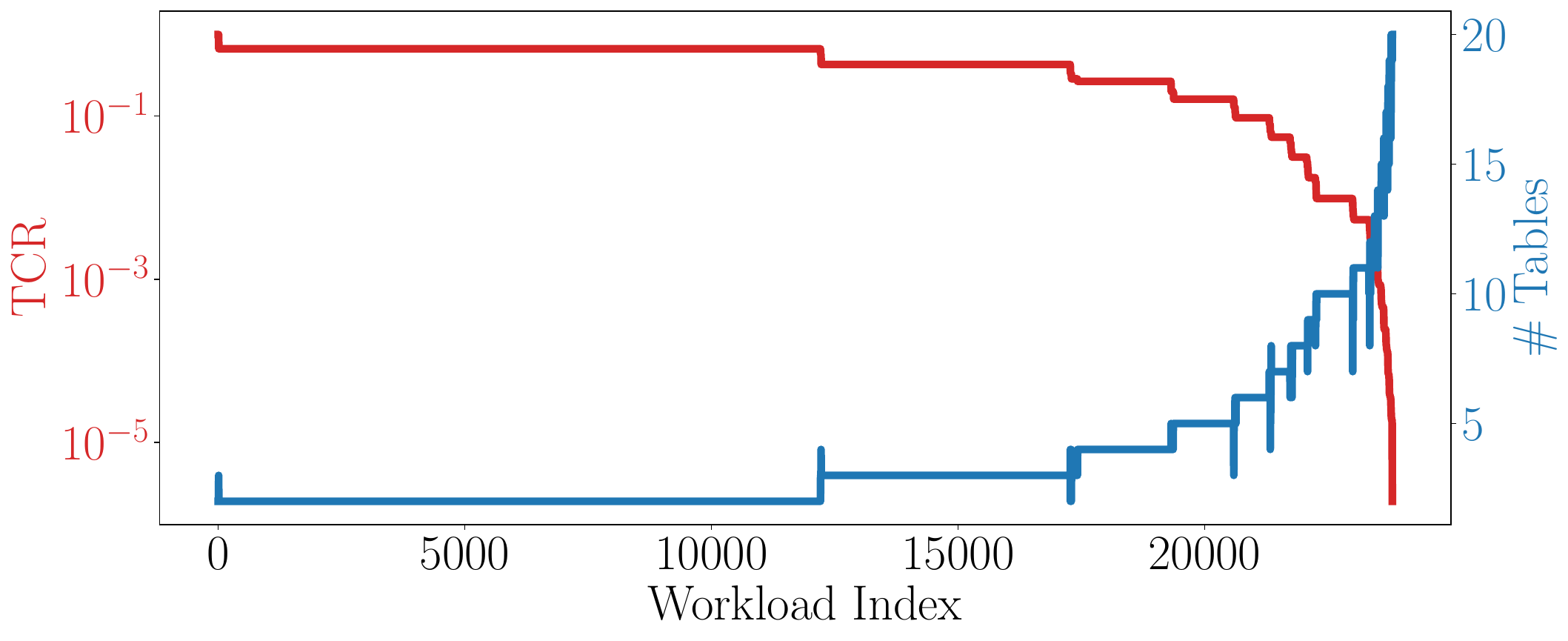}
       \captionsetup{skip=3pt}
    \caption{Join Template Incompleteness}
    \label{subfig:business_imcomplete}
  \end{subfigure}
  \hfill
  \begin{subfigure}{0.23\textwidth}
      \includegraphics[width=\textwidth]{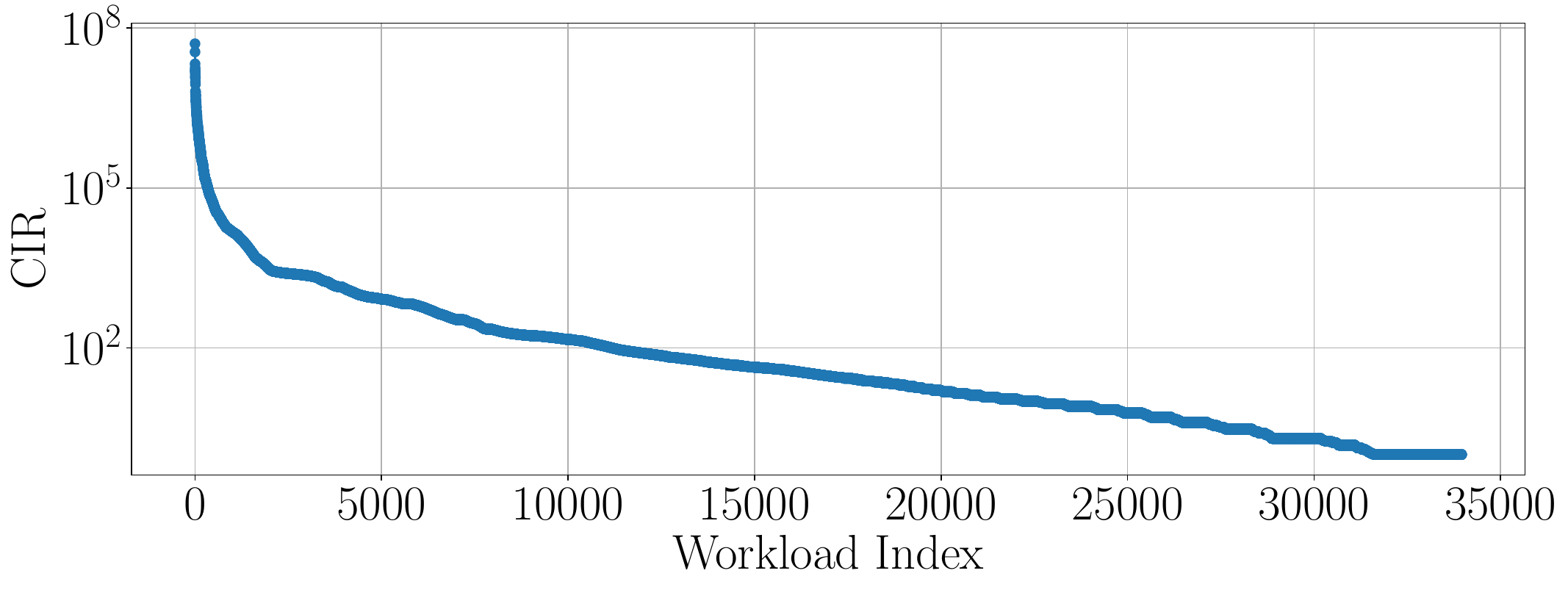}
        \captionsetup{skip=3pt}
      \caption{Join Template Imbalance}
      \label{subfig:business_imbalance}
  \end{subfigure}
  \begin{subfigure}{0.23\textwidth}
    \includegraphics[width=\textwidth]{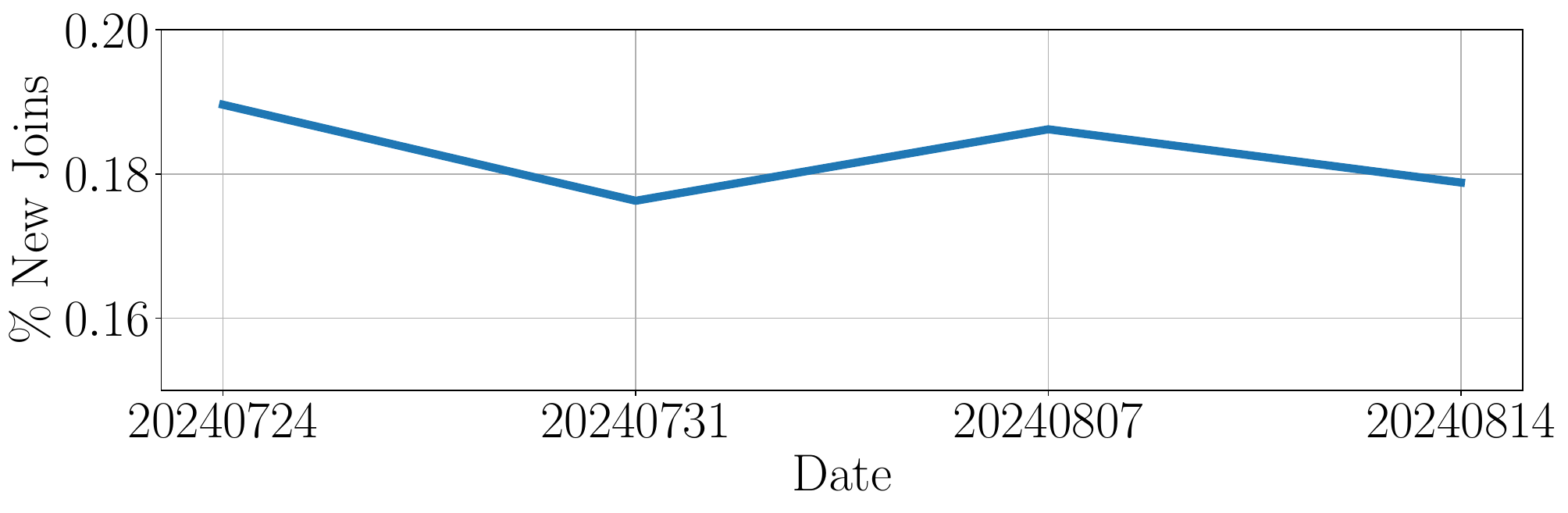}
       \captionsetup{skip=3pt}
    \caption{Join Templates over Time}
    \label{subfig:business_join_shift}
\end{subfigure}
  \hfill
  \begin{subfigure}{0.23\textwidth}
    \includegraphics[width=\textwidth]{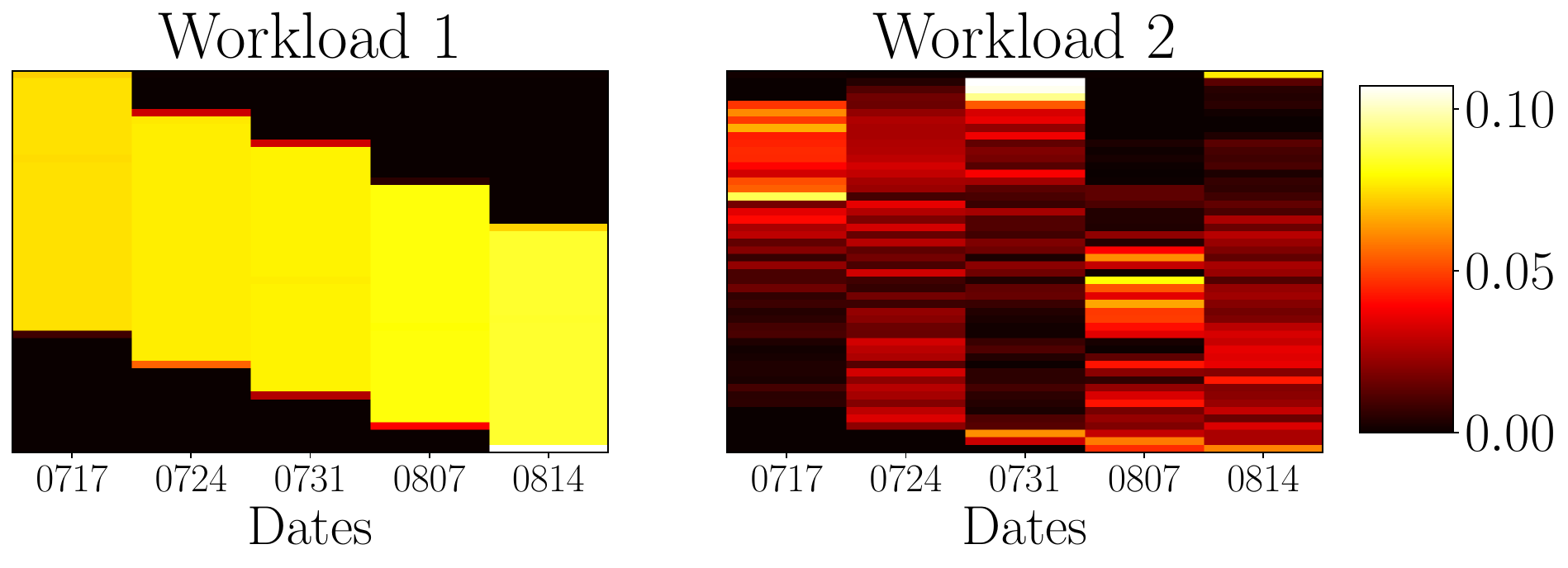}
       \captionsetup{skip=3pt}
    \caption{Value Distribution Shift}
    \label{subfig:business_query_shift}
  \end{subfigure}
  \vspace{-1em}
    \caption{Production Workload Analysis at ByteDance}
  \label{fig:bytedance_workload}
\end{figure}

In Figure~\ref{subfig:business_imcomplete}, the red line represents the TCR of join templates during the period; the blue line represents the number of tables involved in the workload.
Only $\sim 0.03\%$ of workloads' query templates combine all possible table combinations, implying  that \textbf{production workloads seldom contain complete join templates.}

Figure~\ref{subfig:business_imbalance} shows the imbalance in join templates within each workload. We use the Class Imbalance Ratio (CIR) to measure the imbalance of join templates in a workload, defined as the ratio of the most and least common join templates. A higher CIR suggests a greater imbalance. In Figure~\ref{subfig:business_imbalance}, $\sim60\%$ of the workloads have a CIR $>10$, and 20\% have a CIR $>1000$, indicating \textbf{a significant presence of imbalanced workloads in production environments.}

As illustrated in Figure~\ref{subfig:business_join_shift}, each week, each workload exhibits $\sim 18\%$ new query templates on average that were unseen in the previous week. This indicates that \textbf{in business scenarios, the appearance of new join templates is common.}

Figure~\ref{subfig:business_query_shift} uses two business workloads to illustrate weekly variations in predicate values from fixed templates. The x-axis denotes dates, while the y-axis represents normalized value distributions of a predicate. The left shows a systematic shift due to time-increasing IDs focused on the latest week, whereas the right depicts irregular shifts from random user IDs. Both figures highlight that \textbf{predicate values within the same query template can change over time.}


\section{The \name System~\label{grasp}}
In this section, we define the problem, discuss the key design choices of \name, and provide an overview of the \name system.

\subsection{Problem Overview~\label{section.problem_formulation}}


\noindent \textbf{\underline{Problem}:} {\emph{Data-Agnostic Cardinality Learning from Imperfect Query Workloads (DACL)}}.

\noindent  \textbf{\underline{Inputs:}}   We are provided with a set a of SPJ queries $\mathcal{Q}=\{q_i\}_{i=1}^n$ and their ultimate cardinalities $\mathcal{C}=\{c(q_i)\}_{i=1}^n$ 
collected from a database instance $\mathcal{DB}$, along with the database schema for $\mathcal{DB}$ and the table cardinalities (\textit{i.e,} numbers of total rows) $\{|T_j|\}_{j=1}^m$, where $m$ is the number of tables in $\mathcal{DB}$. Training queries do not contain predicates on join keys, {\color{black}since we observe only $2\%$ of real-world queries in production environments apply filters on join keys.}

\noindent  \textbf{\underline{Constraints:}}
\textbf{1). No Data Access:} Direct access to the database \( \mathcal{DB} \) is prohibited. This means no samples, histograms, or direct statistics about the data are available. 
\textbf{2). Incomplete Join Templates:} The training queries do not cover all possible join templates,  leaving most unseen join templates untrained. We use TCR as the measure for incompleteness, \textit{e.g., $\mathrm{TCR}=0.3$}.
 \textbf{3). Imbalanced Join Templates:} The distribution of training queries across different join templates is skewed, with some templates having very few queries available for training. {\color{black}  We use CIR as the measure for imbalance, \textit{e.g., $\mathrm{CIR}=100$}.}
\textbf{4). Value Distribution Shifts:} The value distributions for \emph{range predicates} in incoming queries may differ significantly from those seen during training. 
{\color{black}
We focus on \emph{granularity shifts}, \emph{i.e.,} changes in the distribution of granularity defined in Definition~\ref{definition.granularity}.
}

\noindent  \textbf{\underline{Goals:}} The objective is to develop a model/system that can accurately predict the cardinality (or selectivity) for any incoming query $q$, under the following conditions:
\begin{enumerate} [leftmargin=*]
    \item The query might involve a join template that was not seen during training.
    \item The query may come from a join template that has very few examples in the training set. 
    \item The query may contain predicate values with distributions that differ from the training queries.
\end{enumerate}
{\color{black}
Although our primary focus is on imperfect workloads - motivated by real challenges in Bytedance - a desired model should also apply to \emph{perfect} workloads with complete and balanced join templates. Since we developed \name primarily to address real challenges (which are likely to arise in other enterprises as well), our evaluations focus on imperfect workloads to underscore the practical utility of our approach in real-world settings. Nonetheless, in principle \name requires no modification to handle perfect workloads.

}
\subsection{Key Design Choices~\label{section:grasp:designs}}
Unfortunately, to the best of our knowledge, no existing cardinality estimation approach can effectively achieve the goals simultaneously under the constraints in the DACL problem. To achieve this,
we present \name: \textbf{G}eneralizable, and \textbf{R}obust, data-\textbf{A}gno\textbf{S}tic cardinality \textbf{P}rediction system.  Below, we provide an overview of \name, detailing the desiderate (\textbf{D}) and our solutions.

\begin{itemize} [leftmargin=*]

\item {\textbf{D1: Generalization to unseen join templates despite no access to the underlying data.}}

\textbf{Our Solution:} \name achieves this goal through the notion of \emph{compositionality}. The core idea is that, instead of training a single model to handle all join templates \cite{kipf2019learned} or training separate models for each join template \cite{dutt2020efficiently}, we only learn models for \emph{necessary primitives}. First, we learn query-driven CardEst models only for base relations. Second, without data access, it is challenging to model join correlations. \name borrows the concept of \textit{count sketches} in the literature~\cite{rusu2008sketches,alon1999tracking}. However, instead of existing count sketches that are built from data, \name introduces \textit{learned count sketch (LCS)} models to learn \emph{low-dimensional} count sketches that capture join correlations across base relations, \emph{solely from queries}.
Put it all together, \name allows unseen join templates to be addressed by composing the corresponding primitives. For a taste of the compositional design of \name, consider the simple example:

\begin{exmp}\label{exmp.grasp}
Consider a scenario illustrated in Figure~\ref{fig:joinhandling} where base relations \( A \) and \( B \) are joined using a key \( x \) with a large domain size (e.g., \( |\text{dom}(x)| = 10^6 \)). To compute the cardinality \( c(q^{A \Join B}) \) of the join query \( q^{A \Join B} \), it is necessary to consider not only the cardinalities \( c(q^A) \) and \( c(q^B) \) of the subqueries on base relations \( A \) and \( B \), but also the distribution of the join keys within these subqueries. Let \( \mathbf{f}_{q^A} \) and \( \mathbf{f}_{q^B} \) represent the probability distributions of the join key \( x \) in the two subqueries \( q^A \) and \( q^B \), respectively.
 The cardinality of the $q^{A \Join B}$ can be computed as follows,

\begin{align}
{c(q^{A \Join B})} & = c(q^A) \cdot c(q^B) \cdot (\mathbf{f}_{q^A} \boldsymbol{\cdot} \mathbf{f}_{q^B}) ~\label{eq.grasp.1} \\
   &  \approx \underset{\underset{\fcolorbox{black}{yellow!20}{\footnotesize $\textrm{CardEst}_A$}}{\uparrow}}{c(q^A)} \cdot \underset{\underset{\fcolorbox{black}{yellow!20}{\footnotesize $\textrm{CardEst}_B$}}{\uparrow}}{c(q^B)} \cdot (\underset{\underset{\fcolorbox{black}{blue!10}{\footnotesize $\textrm{LCS}_A$}}{\uparrow}}{\mathbf{v}_{q^A}} \boldsymbol{\cdot} \underset{\underset{\fcolorbox{black}{blue!10}{\footnotesize $\textrm{LCS}_B$}}{\uparrow}}{\mathbf{v}_{q^B}}) ~\label{eq.grasp.2} \\
    & \quad \text{where } \underbrace{\mathbf{f}_{q^A} \boldsymbol{\cdot} \mathbf{f}_{q^B}}_{\textrm{ high-dimensional}} \approx \underbrace{\mathbf{v}_{q^A} \boldsymbol{\cdot} \mathbf{v}_{q^B}}_{\textrm{low-dimensional}}  \notag
\end{align}

where $c(q^A)$ and $c(q^B)$ are computed by per-table CardEst models for $A$ and $B$, respectively. $\mathbf{v}_{q^A}$ and $\mathbf{v}_{q^B}$ are low-dimensional  (\textit{e.g.,} $|\mathbf{v}_{q^A}| = |\mathbf{v}_{q^B}| = 500$) count sketch computed by LCS models for $A$ and $B$. In this computation,
(\ref{eq.grasp.1}) is according to the definition of joins; (\ref{eq.grasp.2}) is based on the assumption that LCS models approximate the scalar dot product (i.e., $\mathbf{f}_{q^A} \boldsymbol{\cdot} \mathbf{f}_{q^B}$) through their learned count sketches for queries $q^A$ and $q^{B}$, thus efficiently capturing join correlations between $A$ and $B$.
\end{exmp}

\item {\textbf{D2: Robustness to join template imbalance despite no access to the underlying data.}}

\textbf{Our Solution:} 
\name naturally achieves this goal through its compositional system design. By leveraging compositionality, \name ensures that model predictions are consistent across different join templates. This allows knowledge learned from the "majority" join templates to be effectively transferred to the "minority" join templates, which addresses the issue of join template imbalance and improves prediction accuracy for underrepresented join templates.

\smallskip
\item \textbf{D3: Robustness to value distribution shifts despite no access to the underlying data. }

\textbf{Our Solution:}  \name builds on the \neucdf framework from \cite{wu2024practical}, which learns models to predict cumulative distribution functions (CDFs) rather than direct query selectivities. This approach enhances out-of-distribution generalization for range predicates by ensuring predictions are induced from signed measures, enforcing the additivity constraint of selectivity/cardinality functions. Additionally, \name improves on \neucdf by introducing a novel CDF prediction model, \arcdf, that addresses a key limitation of the original \neucdf framework.
\end{itemize}

{\color{black} \noindent \textbf{Sample Complexity.}
Compared to non-DNN models~\cite{chen1994adaptive, park2020quicksel,stillger2001leo}, the use of DNNs in \name improves the  prediction accuracy due to higher model capacity~\cite{kipf2019estimating, wu2024practical}, at the cost of increasing the need for training queries (\textit{e.g.,} higher sample complexity). However, the sample complexity remains modest. For instance, in our experiments, the maximum number of queries used (on CEB-IMDb-full) is  $\sim 260$K, significantly lower than the 2 million queries generated weekly per workload at ByteDance. The sample complexity of \name and compared models will be further evaluated in \S~\ref{section:complexity}. However, the sample complexity is evaluated  based on the query type outlined in Definition~\ref{def:query}, and introducing additional features such as \texttt{Group-By} would increase the sample complexity, which is a limitation of this work.
}

\subsection{\name Overview~\label{section.system.overview}}
Recall that \name does not have direct access to DBMS data, but it does have access to schema information and a query workload with ultimate cardinality outcomes.

\begin{figure}
\centering
\includegraphics[height=0.16\textwidth]{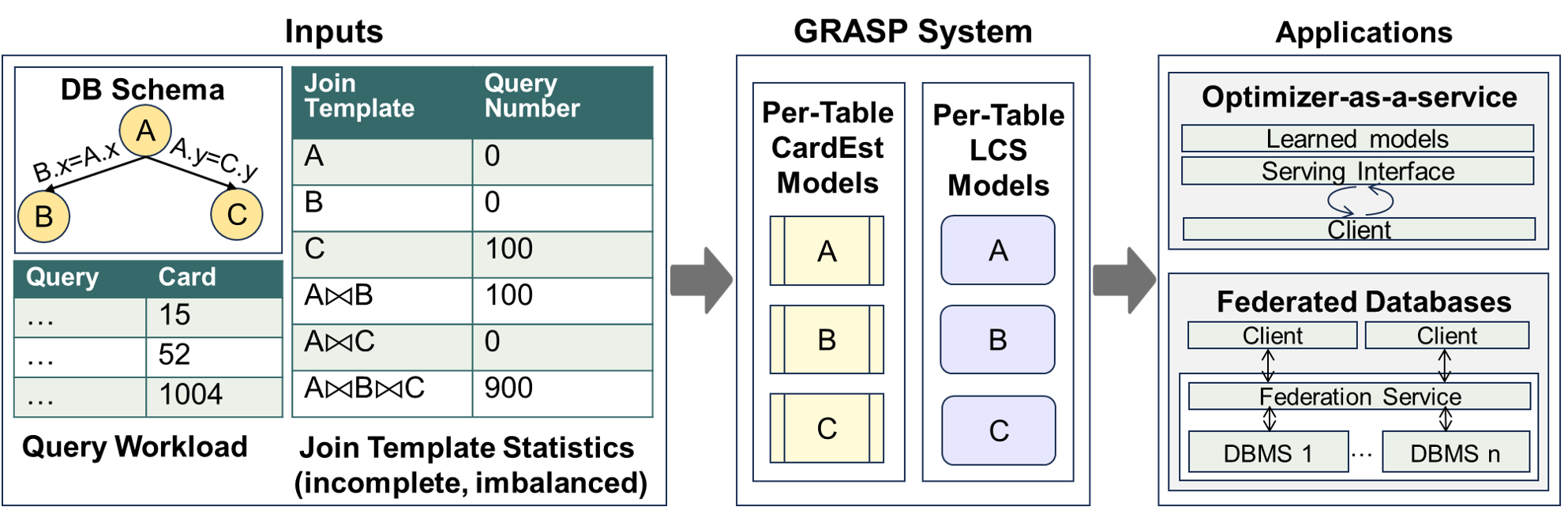}
\caption{Workflow of \name.}
\label{fig:overview}
\end{figure}

\noindent \textbf{\name Workflow.} Combining the design choices introduced before, \name operates in two main stages: the \emph{building stage} and the \emph{serving stage}. The workflow of \name is depicted in Figure~\ref{fig:overview}.
In the building stage, \name ingests the schema of the DB instance along with a collected query workload, which may be incomplete or imbalanced. The goal of this stage is to train the primitive models (\textit{i.e.,} the CardEst model and the LCS model for each base relation) in \name using the query workload. Since all computational steps in CardEst with \name (will be introduced in \S~\ref{section.compo.cardest}) are fully differentiable, we can use modern stochastic optimization methods, such as backpropagation and stochastic gradient descent (SGD), to efficiently learn the model parameters $\theta$ of all primitive models. {\color{black}The primitive models (both CardEst and LCS models) for each base relation $T$ will be trained on queries in which $T$ is involved.} Specifically, training is performed by minimizing the following loss function over the training workload ($N$ training queries):
\begin{equation}\label{eq.lossfuc}
    \mathcal{L} = \Bigl(\sum_{i=1}^{N}\mathrm{Error}\bigl({\hat{c}(q_i)}, c(q_i)\bigl)\Bigl)/N,
\end{equation}
where $\mathrm{Error}$ is the error function measuring the discrepancy between the estimated cardinality $\hat{c}(q_i)$ from \name and the true cardinality $c(q)$. Users can specify this error function, with common choices including Square Error. \textit{i.e.,} SE = $(\hat c(q) - c(q))^2$, Qerror: ($\max{\frac{\hat c(q)}{c(q)}, \frac{c(q)}{\hat c(q)}}$) or Squared Logarithmic Error, \textit{i.e.,} SLE = $(\log{\hat c(q)} - \log{c(q)})^2$. Optimizing SLE is equivalent to optimizing Qerror~\cite{negi2021flow} {\color{black}(resulting in consistent outcomes between SLE and Qerror measurements)}. While \name is compatible with any error functions, we adopt SLE as it yields the best overall performance.

Once built, \name\ can serve as the underlying cardinality estimator for federated databases~\cite{giannakouris2022building}, services that abstract over alternative DBMS platforms~\cite{gadepally2016bigdawg}, and for query optimization-as-a-service~\cite{jindal2022query}. Note that all these applications may not have access to the underlying data. 

\smallskip

\noindent \textbf{Optimizing \name.}
Queries from different join templates may require distinct sets of primitive models. This enables the application of \emph{distributed training} with multiple GPUs, where join templates with non-overlapping primitive models can be trained concurrently across multiple computing nodes. By partitioning the training tasks based on their \emph{primitive model dependencies}, we can assign each group of non-overlapping join templates to separate workers by using a \emph{greedy packing algorithm}~\cite{coffman1999bin} for scheduling. Due to space constraints, we omit the details in this paper.

To perform query optimization for a join query, we need to estimate the cardinalities of all subqueries. Leveraging a GPU can accelerate this process through batched inference. However, even with CPU-based inference, \emph{progressive query inference} can significantly reduce \name's computational overhead, due to its compositional design. This approach estimates subqueries in a bottom-up fashion, starting with those containing fewer joins. By saving and reusing results from these smaller joins, we avoid redundant computations in estimations of larger joins, enhancing overall efficiency.

\section{\mbox{Compositional Generalization}~\label{section.compo_design}}

\subsection{Motivating Examples~\label{section.compo_design.motivation}}

Consider the database instance in figure~\ref{fig.motivation} with three base relations $\{A, B, D\}$, each of cardinality 5. They can be joined via join keys $A.x, B.x, D.x$ (with no restrictions on primary or foreign keys), which belong to a join group $x$ with domain: $\text{dom}(x)=\{1,2\}$. Figure~\ref{fig.motivation.b} shows four possible join templates: $A \Join B$, $A \Join D$, $B \Join D$, and $A \Join B \Join D$.

\noindent \textbf{Inferring cardinalities for queries over a ``hidden'' table.} 
We begin by examining the scenario where a specific table is absent from any single-table queries within the training workload. Instead, the workload comprises single-table queries on other tables and join queries that involve this ``hidden'' table. Our objective is to explore the feasibility of inferring cardinality estimates for the hidden table under these conditions.

We first focus on deriving the \textcolor{palebrown}{unknown} cardinalities of queries $q^{B}$ on relation $B$ by leveraging  \textcolor{darkgreen}{known} cardinalities from queries $q^{A}$ on relation $A$ and queries $q^{A \Join B}$ on join template $A \Join B$. Similar to Example~\ref{exmp.grasp}, calculating the cardinality of a join query $q^{A \Join B}$ needs the cardinalities of the individual subqueries $q^{A}, q^{B}$ on base relations $A, B$, and the distributions of join keys in both. Here we simplify the problem by assuming a small domain size for $x$, $|\text{dom}(x)|=2$.
Specifically, assume that the join key distributions for $x \in \{1,2\}$ in $q^A$, $q^B$ 
 are $\mathbf{p}^{A}=[p^{A}_1,p^{A}_2]^\top,  \mathbf{p}^{B}=[p^{B}_1,p^{B}_2]^\top$. Let $\mathbf{L}= [c(q^A), c(q^B)]$ and $\mathbf{P}=[\mathbf{p}^{A}, \mathbf{p}^{B}]^\top$, by definition of joins, the cardinality  of the $q^{A \Join B}$ is 
\begin{equation}
   \textcolor{darkgreen}{c(q^{A \Join B})} =  \Vert \mathbf{L} \mathbf{P} \Vert_1  = \underbrace{\textcolor{darkgreen}{c(q^{A})} \textcolor{palebrown}{p^{A}_1}   \textcolor{palebrown}{c(q^{B})} \textcolor{palebrown}{p^{B}_1} }_{\text{join key } x=1} + \underbrace{\textcolor{darkgreen}{c(q^{A})}  \textcolor{palebrown}{p^{A}_2}  \textcolor{palebrown}{c(q^{B})} \textcolor{palebrown}{p^{B}_2}}_{\text{join key } x=2} . \label{eq.example.joincard}
\end{equation}
Eq.~\ref{eq.example.joincard} can be easily generalized to joins (via $x$) with $>2$ relations.

Apparently, given the join key distributions of queries from tables $A$ and $B$  (\textit{i.e.,} $\mathbf{P}$), we can estimate the cardinality \(\textcolor{palebrown}{c(q^{B})}\) of any query \( q^{B} \) using Eq.~\ref{eq.example.joincard}. However, these distributions are typically \emph{hidden} and not directly available. Therefore, in order to infer \(\textcolor{palebrown}{c(q^{B})}\), we also need to infer these join key distributions from known cardinalities, \(\textcolor{darkgreen}{c(q^{A})}\) and \(\textcolor{darkgreen}{c(q^{A \Join B})}\). This approach is viable provided we have a sufficient number of queries for \( q^{A} \) and \( q^{A \Join B} \).

\vspace{-0.5em}
\begin{figure}[ht]
\subcaptionbox{Base relations ($\{A, B, D\}$)}[0.28\textwidth]{
\begin{tikzpicture}
    \matrix(dict)[matrix of nodes,ampersand replacement=\&,
        nodes={align=center,text width=1.4em},
        is row /.style={anchor=south},
         column /.style={nodes={text width=0.9em,align=left}}
    ]{
        $A.x$ \& $A.a$  \\
        1 \& 1  \\
        2 \& 2 \\
        1 \& 3 \\
        1 \& 4 \\
        2 \& 5 \\
    };\quad \quad \quad
   \draw(dict-1-1.south west)--(dict-1-2.south east);
   \draw(dict-1-1.north east)--(dict-6-1.south east);
    
\end{tikzpicture}
\begin{tikzpicture}
   \matrix(dict)[matrix of nodes,ampersand replacement=\&,
        nodes={align=center,text width=1.4em},
        is row /.style={anchor=south},
         column /.style={nodes={text width=0.2em,align=left}}
    ]{
        $B.x$ \& $B.b$  \\
        1 \& 2  \\
        1  \& 2 \\
        2  \& 3 \\
         1  \& 4 \\
         2  \& 4 \\
    };\quad \quad \quad
   \draw(dict-1-1.south west)--(dict-1-2.south east);
   \draw(dict-1-1.north east)--(dict-6-1.south east);
\end{tikzpicture}
\begin{tikzpicture}
    \matrix(dict)[matrix of nodes,ampersand replacement=\&,
        nodes={align=center,text width=1.4em},
        is row /.style={anchor=south},
    ]{
        $D.x$ \& $D.d$  \\
        1 \& 4  \\
        1 \& 5 \\
        1 \& 5 \\
        2 \& 6 \\
        2 \& 7 \\
    };
  \draw(dict-1-1.south west)--(dict-1-2.south east);
  \draw(dict-1-1.north east)--(dict-6-1.south east);
 \end{tikzpicture}
}
\subcaptionbox{All join templates \label{fig.motivation.b}}[0.17 \textwidth]{
 \begin{tikzpicture}[
roundnode/.style={circle, draw=black, fill=white, thick, scale=0.7},
]
\node[roundnode]   (a)     {$A$};
\node[roundnode]        (b)       [right=1.1cm of a] {$B$};

\draw[-, thick] (b.west) -- node[above]{\scriptsize $A.x = B.x$} (a.east);
\end{tikzpicture}
\begin{tikzpicture}[
roundnode/.style={circle, draw=black, fill=white, thick, scale=0.7},
]
\node[roundnode]   (a)     {$A$};
\node[roundnode]        (b)       [right=1.1cm of a] {$D$};

\draw[-, thick] (b.west) -- node[above]{\scriptsize $A.x = D.x$} (a.east);
\end{tikzpicture}
\begin{tikzpicture}[
roundnode/.style={circle, draw=black, fill=white, thick, scale=0.7},
]
\node[roundnode]        (b)       [right=1.1cm of a] {$B$};
\node[roundnode]      (c)       [right=1.1cm of b] {$D$};

\draw[-, thick] (b.east) -- node[above]{\scriptsize $B.x = D.x$} (c.west);
\end{tikzpicture}

\begin{tikzpicture}[
roundnode/.style={circle, draw=black, fill=white, thick, scale=0.7},
]
\node[roundnode]   (a)     {$A$};
\node[roundnode]        (b)       [right=1cm of a] {$B$};
\node[roundnode]      (c)       [right=1cm of b] {$D$};

\draw[-, thick] (b.west) -- node[above, font=\scriptsize]{$A.x = B.x$} (a.east);
\draw[-, thick] (b.east) -- node[above]{\scriptsize $B.x = D.x$} (c.west);
\end{tikzpicture}  
}
\vspace{-0.5em}
\caption{The database instance for motivating examples.} \label{fig.motivation}
\end{figure}

\vspace{-1.5em}
\begin{exmp}\label{exmp.infer}
\begin{table}[ht]
    \centering
    \begin{minipage}{0.21\textwidth}
        \caption{Queries on table $A$} \label{tab:motivation:stbs}
    \renewcommand{\arraystretch}{1.2} 
    \vspace{-4mm}
        \centering
        \scalebox{0.79}{
        \begin{tabular}{|c|c|c|}
            \hline
            \textbf{Query} & \textbf{Predicate} & \textbf{Card} \\
            \hline
            \( q^{A}_1 \) & \( \{ a\leq2 \} \) & 2 \\
            \hline
            \( q^{A}_2 \) & \( \{ 2<a\leq4 \} \) & 2 \\
            \hline
            \( q^{A}_3 \) & \( \{  a > 4 \} \) & 1 \\
            \hline
        \end{tabular}}
        \vspace{+1em}
        \caption{Queries on table $B$} \label{tab:motivation:stbsB}
    \vspace{-4mm}
        \scalebox{0.79}{
        \begin{tabular}{|c|c|c|}
            \hline
            \textbf{Query} & \textbf{Predicate} & \textbf{Card} \\
            \hline
            \( q^{B}_1 \) & \( \{ b\leq3 \} \) & \textcolor{palebrown}{?} \\
            \hline
            \( q^{B}_2 \) & \( \{ b>4 \} \) & $5-\textcolor{palebrown}{c(q^{B}_1)}$ \\
            \hline
        \end{tabular}}
    \end{minipage}
    \begin{minipage}{0.26\textwidth}
        \caption{Queries on join \( A \Join B \)}
        \label{tab:motivation:joins}
    \vspace{-4mm}
     \renewcommand{\arraystretch}{1.3} 
        \centering
        \scalebox{0.79}{
        \begin{tabular}{|c|c|c|}
        \hline
        \textbf{Query} & \textbf{Predicate} & \textbf{Card} \\
        \hline
        \( q^{A \Join B}_1 \) & \( \{ a \leq 2 \wedge b \leq 3 \} \) & 3 \\
        \hline
        \( q^{A \Join B}_2 \) & \( \{ a \leq 2 \wedge b > 3 \} \) & 2 \\
        \hline
        \( q^{A \Join B}_3 \) & \( \{ 2 < a \leq 4 \wedge b \leq 3 \} \) & 4 \\
        \hline
        \( q^{A \Join B}_4 \) & \( \{ 2 < a \leq 4 \wedge b > 3 \} \) & 2 \\
        \hline
        \( q^{A \Join B}_5 \) & \( \{ a > 4 \wedge b \leq 3 \} \) & 1 \\
        \hline
        \( q^{A \Join B}_6 \) & \( \{ a > 4 \wedge b > 3 \} \) & 1 \\
        \hline
    \end{tabular}}
    \end{minipage}
\end{table}

In our query workload, we have three queries on relation \(A\) (as listed in Table~\ref{tab:motivation:stbs}) and six join queries on \(A \Join B\) (detailed in Table~\ref{tab:motivation:joins}). Each query is represented by a tuple consisting of (the query name, predicate, and cardinality). For notation, we define \(p^T_{i}\) as the proportion of tuples in the query \(q_i^{T}\) on relation \(T\) with join key \(x=1\); thus, the remaining proportion with \(x=2\) is \(1-p^T_{i}\). For instance, \(p^A_{1}\) denotes the proportion of tuples in the query \(q_1^A\) on \(A\) with join key \(x=1\).
Our objective is to deduce the cardinalities of queries on the ``hidden'' relation \(B\), specifically for queries \(q_1^B\) and \(q_2^B\) as outlined in Table~\ref{tab:motivation:stbsB}. Thus, the primary unknown is \(\textcolor{palebrown}{c(q^{B}_1)}\)

The challenge lies in the absence of join key distributions for queries on \(A\) and \(B\), which include five unknowns: \(\textcolor{palebrown}{p^A_{1}}\), \(\textcolor{palebrown}{p^A_{2}}\), \(\textcolor{palebrown}{p^A_{3}}\), \(\textcolor{palebrown}{p^B_{1}}\), and \(\textcolor{palebrown}{p^B_{2}}\). However, the six join queries from \(q_1^{A \Join B}\) to \(q_6^{A \Join B}\) provide a basis to establish six linearly dependent equations using Eq.~\ref{eq.example.joincard}.
Combining these with the three known cardinalities from queries on \(A\) (i.e., \(\textcolor{darkgreen}{c(q^{A}_1)}\), \(\textcolor{darkgreen}{c(q^{A}_2)}\), \(\textcolor{darkgreen}{c(q^{A}_3)}\)) and employing linear algebra techniques, we can solve these equations to determine \(\textcolor{palebrown}{c(q^{B}_1)}=3\) along with other unknowns. 

\end{exmp}

Note that the approach of ``decomposing large queries into smaller queries'' applies to queries of any size (\textit{e.g.,} queries with more than two joins).  In other words, small join templates (e.g., involving hidden tables) can be effectively learned from the larger join templates they are part of.

\noindent \textbf{Composing a novel join template.} Now, we explore composing queries of a novel join template that was unseen in the workload. We use information from the query workload, as well as inferred query cardinalities and join correlations. Before moving forward, we note that Eq.~\ref{eq.example.joincard} can be generalized to accommodate join templates involving multiple tables (i.e., number of tables $n_d>2$) that belong to the same join group, say $x$. In this generalized case, $\mathbf{L}$ is a vector of length $n_d$, representing the cardinalities of single-table subqueries on each base relation. Additionally, $\mathbf{P}$ is an $n_d \times n_k$ matrix, where $n_k$ is the number of distinct values of join key $x$. Each row of $\mathbf{P}$ is the join key distribution for the corresponding single-table subquery.

\begin{exmp} \label{exmp.compose}
Assume that apart from the queries discussed in Example~\ref{exmp.infer}, the query workload includes an additional query on table \(D\), \(q_1^D:\{d \leq 5\}\) with \(\textcolor{darkgreen}{c(q_1^D) = 3}\); and one query on the join template \(B \Join D\), \(q_1^{B \Join D}:\{b \leq 3 \wedge d \leq 5\}\) with \(\textcolor{darkgreen}{c(q_1^{B \Join D}) = 6}\). The goal is now to compose queries on a larger join template \(A \Join B \Join D\) that has not been observed in the query workload. 

From Example~\ref{exmp.infer}, we already determined that \(\textcolor{palebrown}{c(q^B_{1})}=3\) and \(\textcolor{palebrown}{p^B_{1}} = \frac{2}{3}\). Therefore, we can establish the following equation for \(q_1^{B \Join D}\) with only one unknown, \(\textcolor{palebrown}{p^D_{1}}\), using Eq.~\ref{eq.example.joincard}:

\begin{equation*}
    3\cdot\frac{2}{3}\cdot3\cdot \textcolor{palebrown}{p^D_{1}} +  3\cdot(1-\frac{2}{3})\cdot3\cdot(1-\textcolor{palebrown}{p^D_{1}}) = 6
\end{equation*}
Solving this equation, we have $\textcolor{palebrown}{p^D_{1}} = 1$. Then we can derive the cardinality of the query $q_1^{A \Join B \Join C}: \{a\leq2 \wedge b\leq3 \wedge d\leq5\}$ by using the generalized version of Eq.~\ref{eq.example.joincard}:
\begin{equation*}
c(q_1^{A \Join B \Join C}) =  \underbrace{2\cdot\frac{1}{2}\cdot3\cdot\frac{2}{3}\cdot3\cdot1}_{\text{join key } x=1} + \underbrace{2\cdot\frac{1}{2}\cdot3\cdot\frac{1}{3}\cdot3\cdot 0}_{\textrm{join key }x=2}  = 6
\end{equation*}

\noindent Other queries such as $q_2^{A \Join B \Join C}: \{2<a\leq4 \wedge b\leq3 \wedge d\leq5\}$ and $q_3^{A \Join B \Join C}: \{a>4 \wedge b\leq3 \wedge d\leq5\}$ can be composed similarly as we have derived necessary elements (\textit{e.g.,} cardinalties of single-table subqueries and corresponding join key distributions).
\end{exmp}

\subsection{The Compositional Design~\label{section:compo:design}}
From the previous examples, we draw two key takeaways. \textbf{First}, smaller join templates not present in the query workload could be inferred from larger ones. 
\textbf{Second}, by combining explicit workload information (\textit{e.g.,} cardinalities of training queries) with inferred information (\textit{e.g.,} cardinalities of hidden relations and join key distributions), we can reliably estimate the cardinalities of queries from unseen join templates. 
Therefore, a key question arises: \emph{how can we more effectively extract and utilize both the explicit and implicit information from the query workload to perform accurate cardinality estimation for as many join templates as possible?} This informs the compositional design of \name.

\smallskip

\noindent \textbf{Compositionality.} Compositionality as compositional generalization, refers to the capacity for systematic generalization to new, combined examples from a specific distribution, after training on a different distribution that introduced the necessary \emph{components/primitives}. This concept emphasizes the ability to understand and apply combinations of learned primitives in novel contexts. For example, Pavel et.al~\cite{tokmakov2019learning} trains an image classification model that decomposes concepts into parts and allows it to generalize to novel categories with fewer examples.

\smallskip

\noindent \textbf {Primitives.}
With regard to the DACL problem, we introduce two types of primitive models for each base relation:  1) Per-table CardEst models and 2) Per-table Join Key (JK) models, both receiving single-table queries on the associated relation as inputs. Per-table CardEst models predict the cardinality of the query, whereas JK models output the distribution of the join key in the query results (thus capturing the correlations among different base relations via join keys). If a base relation includes multiple join keys, the JK models output a set of distributions, one for each join key.
We formally define their abstracts as follows.
\begin{itemize} [leftmargin=*]
    \item \textbf{Per-table CardEst Models $\mathcal{M}_{CE}$: }  \(\text{ Query} \to \text{Cardinality} \)

    \item \textbf{Join Key (JK) Models  $\mathcal{M}_{JK}$: } \(\text{Query} \to \text{\{Join Key Distribution\}} \)
\end{itemize}

 \name builds these primitive models from queries, enabling generalizable and consistent cardinality estimation across join templates, which is the key to achieving \textbf{D1} and \textbf{D2} in \S~\ref{section.system.overview}. Since these primitive models can be effectively trained on any relevant queries, they do not require specific single-table or two-table queries.

The use of join key models, \(\mathcal{M}_{JK}\), has exhibited certain limitations. Consequently, we will revisit and revise the design of \(\mathcal{M}_{JK}\) by replacing it with the \emph{learned count sketch} models, \(\mathcal{M}_{LCS}\), as detailed in \S~\ref{section.alv}. For the purpose of our current discussion, we will provisionally consider \(\mathcal{M}_{JK}\) and \(\mathcal{M}_{LCS}\) to be interchangeable.

\noindent \textbf{Model Choices for \(\mathcal{M}_{CE}\).} Note that the compositional design of \name\ is general, which means any query-driven model architectures can be used for \(\mathcal{M}_{CE}\). For tables involving range predicates, \name uses the ArCDF model as it is more robust to value distribution shifts for range predicates (will be introduced in \S~\ref{section.arcdf}). For tables involving other types of predicates (such as \texttt{LIKE}, \texttt{IN}, and more), \name employs the Multi-Set Convolutional Networks (MSCN)~\cite{kipf2019learned} for \(\mathcal{M}_{CE}\), as the MSCN encoding mechanism is flexible enough to incorporate complex operators (\textit{e.g.}, \texttt{LIKE})~\cite{negi2021flow}. For details about the MSCN encoding, refer to ~\cite{kipf2019learned}. We will discuss the model choices for $\mathcal{M}_{LCS}$ in \S~\ref{section.alv}.

\subsection {Cardinality Estimation with Primitives~\label{section.compo.cardest}}
To estimate the cardinality $\hat{c}(q)$ of a query $q$, we use the CardEst models $\{\mathcal{M}_{CE}\}$ and join key models $\{\mathcal{M}_{JK}\}$ via \name's main function: $\hat{c}(q) = \texttt{Est}(q, \{\mathcal{M}_{CE}\}, \{\mathcal{M}_{JK}\})$.

\subsubsection {\textbf{Single-table Queries}} Answering queries over base relations is trivial --- \name uses the corresponding per-table CardEst models to estimate the cardinality results. 

\subsubsection {\textbf{Join Queries}} Estimating join queries involves the use of both per-table CardEst models and join key models that are related to the join queries. First, we define an \emph{equivalence join key group} as the set of keys that are joinable according to the database schema. For example, if \(A.x = B.x\) and \(B.x = D.x\) are specified in the schema, then \(A.x\), \(B.x\), and \(D.x\) are considered to belong to the same join key group \(x\). The full set of join key groups and the table ordering within each key group can be easily precomputed from the database schema~\cite{kim2024asm}. Now, we begin with an easier case of join estimation.

\noindent \textbf{Case 1: One join key group.} Consider a join query $q$ that involve a set $\mathcal{T}$ of base relations. They are joined via join keys in the same join key group. Similar to Example~\ref{exmp.compose}. \name employs four steps to calculate the query cardinality $\hat{c} $. 

\begin{enumerate}
    \item Calculating the cardinality estimates for single-table subqueries (\textit{i.e.,} $\hat{c}(q^T)$ for $T\in \mathcal{T}$), using associated $\mathcal{M}_{CE}$.
    \item Calculating the join key distributions (\textit{i.e.,} $\hat{f}(q^T)$ for $T \in \mathcal{T}$) for single-table subqueries, using associated $\mathcal{M}_{JK}$.
    \item Calculating the frequencies of join keys in each single-table query: $\widehat{freq}(q^T) = \hat{c}(q^T)\cdot\hat{f}(q^T)$ for $T \in \mathcal{T}$.
     \item Multiplying the frequencies for every single query and then summing over the result: $\hat{c} = \big\Vert\prod_{T \in \mathcal{T}}{\widehat{freq}(q^T)}\big\Vert_1$.
\end{enumerate}

\begin{algorithm}
    \centering
    \caption{Estimating a join query with \name}\label{alg.cardset}
    \begin{algorithmic}[1]
        \REQUIRE Per-table CardEst models $\{ \mathcal{M}_{CE}\}$ and JK models $\{ \mathcal{M}_{JK}\}$; Query $q$ with involved tables $\mathcal{T}$ and join key groups $\mathcal{G} = \{G\}$.
        \ENSURE \name's cardinality estimate of $q$: $\hat{c}(q)$.
        
         \STATE \texttt{\# compute cardinalities for per-table subqueries}
         \FOR{$T \in \mathcal{T}$} 
          \STATE $\hat{c}(q^T) = \mathcal{M}^T_{CE}(q^T)$ \COMMENT{\texttt{associated CardEst model}}
        \ENDFOR
        
        \STATE \texttt{\# computing join estimates}
         \STATE $\mathcal{T}_{curr} \leftarrow \{\}$, $\hat{c} \leftarrow 1$
        \WHILE{$\mathcal{G}$ is not empty} 
             \STATE $G= \text{FindNextGroup}(\mathcal{T}_{curr}, \mathcal{G})$
          \STATE $\text{key}_G \leftarrow \text{join key of the group $G$}$
                  \STATE $\hat{f} \leftarrow \mathbf{1}_{|\text{dom}(\text{key}_G)|}$ \COMMENT{\texttt{Init. distrib. vector}}
          \FOR{$T \in G.\text{tables}$} 
            \STATE $\hat{f} = \hat{f} \odot \mathcal{M}^{T}_{JK}(q^T, \text{key}_G)$ \COMMENT{\texttt{associated JK model}}
            \IF{\textbf{not} $T \in \mathcal{T}_{curr}$}
                \STATE $\hat{c} = \hat{c} * \hat{c}(q^T)$
                \STATE add $T$ to $\mathcal{T}_{curr}$
            \ENDIF 
          \ENDFOR
           \STATE $\hat{c} =  \hat{c}*\Vert \hat{f} \Vert_1$
           \STATE Remove $G$ from $\mathcal{G}$
        \ENDWHILE
    \RETURN $\hat{c}$

    \end{algorithmic}
\end{algorithm}

\noindent \textbf{Case 2: Multiple join key groups.} \name can be extended to handle the case when the join query $q$ contains join keys from a set of multiple join key groups, $\mathcal{G} = \{G\}$. The estimation procedure is shown in Algorithm~\ref{alg.cardset}.

The algorithm estimates the cardinality of a $q$ by leveraging both per-table CardEst models $\mathcal{M}_{CE}$ and join key models $\mathcal{M}_{JK}$. It begins by calculating cardinality estimates for single-table queries on each relation using corresponding $\mathcal{M}_{CE}$ (lines 1-3). Then, it recursively processes groups of join keys by finding joinable groups (the implementation of FindNextGroup is straightforward using the database schema and can be pre-computed) that connect tables already considered (lines 19-27). For each group, it initializes a distribution vector (lines 9) and iteratively refines it by multiplying with the join key distribution estimate from the associated JK model $\mathcal{M}_{JK}$ (lines 10-11, where $\mathcal{M}^{T}_{JK}(q^T, \text{key}_G)$ represents the predicted distribution of $\text{key}_G$ in query $q^T$ by model $\mathcal{M}^{T}_{JK}$). The overall cardinality estimate is updated by multiplying with the sizes of newly included tables and the \emph{sum} of the refined distribution vector (lines 12-15). This process continues until all join key groups have been incorporated, ultimately yielding the estimated cardinality of $q$.

\subsubsection {\textbf{Remark}}
Acute readers might notice that Algorithm~\ref{alg.cardset} assumes that join keys \emph{within a table} are independent. Despite this simplification, it works well in practice without requiring much computation. Although \name\ could be naturally extended to account for these correlations, our tests show that this only slightly improves accuracy while significantly increasing computational costs. Therefore, \name\ keeps the independence assumption to balance efficiency and accuracy for real-world use.

\section{Building Robust CardEst Models~\label{section.arcdf}}Recall that in \textbf{D3} of \S~\ref{section.system.overview}, we outline the key goal for per-table CardEst models: robustness to value distribution shifts for range predicates. This section presents our solution --- \arcdf.

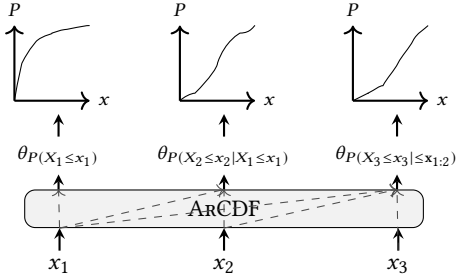
\begin{figure}
\centering
\scalebox{1}{
\begin{tikzpicture}

    \tikzstyle{nn} = [rectangle, rounded corners, minimum width=3cm, minimum height=0.3cm,text centered, draw=black, fill=gray!10]
    \tikzstyle{arrow} = [thick,->,>=stealth]
    \tikzstyle{spline} = [draw, thick, color=blue]

    \node (X1) {$x_1$};
    \node (X2) [right=1.67cm of X1] {$x_2$};  
    \node (Xd) [right=1.8cm of X2] {$x_3$};  

    \node (ARModel) [above=0.3cm of X2, nn, minimum width=5.3cm,  minimum height=0.5cm] {\textsc{ArCDF}};  

    \node (CDF1) [above=0.2cm  of ARModel, xshift=-2.2cm] {\footnotesize $\theta_{P(X_1 \leq x_1)}$};
    \node (CDF2) [above=0.2cm  of ARModel] {\footnotesize $\theta_{P(X_2 \leq x_2 | X_1 \leq x_1)}$};
    \node (CDFd) [above=0.2cm  of ARModel, xshift=2.3cm] {\footnotesize $\theta_{P(X_3 \leq x_3 | \leq \mathbf{x}_{1:2})}$};

  
    \node (NN1) [above=0.3cm of X1, draw=none]{};
    \draw[arrow] (X1) -- (NN1);
     \draw[arrow] (X2) -- (ARModel);
    \node (NNd) [above=0.3cm of Xd, draw=none]{};
    \draw[arrow] (Xd) -- (NNd);

     \node (armodetmp1) [below=0.2cm of CDF1, draw=none]{};
    \draw[arrow] (armodetmp1) -- (CDF1);
    
    \draw[arrow] (ARModel.north) -- (CDF2.south);

    \node (armodetmp2) [below=0.2cm of CDFd, draw=none]{};
    \draw[arrow] (armodetmp2) -- (CDFd);

    \draw[->, dashed, color=black!70] (NN1.south) -- (armodetmp1.north);
    \draw[->, dashed, color=black!70] (NN1.south) -- (ARModel.north);
    \draw[->, dashed, color=black!70] (NN1.south) -- (armodetmp2.north);

    \draw[->, dashed, color=black!70] (ARModel.south) -- (ARModel.north);
    \draw[->, dashed, color=black!70] (ARModel.south) -- (armodetmp2.north);

    \draw[->, dashed, color=black!70] (Xd) -- (armodetmp2.north);

    \node (drawtmp1) [above=0.3cm of CDF1, draw=none]{};
    \draw[arrow] (CDF1) -- (drawtmp1);

     \node (drawtmp2) [above=0.3cm of CDF2, draw=none]{};
    \draw[arrow] (CDF2) -- (drawtmp2);

     \node (drawtmpd) [above=0.3cm of CDFd, draw=none]{};
    \draw[arrow] (CDFd) -- (drawtmpd);

    \begin{scope}[shift={(-0.6, 2.2cm)},  scale=1]
    \coordinate (A) at (0,0);
        \coordinate (A) at (0,0);
        \coordinate (B) at (0.1,0.5);
        \coordinate (C) at (0.3,0.8);
        \coordinate (D) at (0.5,0.9);
        \coordinate (E) at (0.7,0.95);
        \coordinate (F) at (1,1);
        \draw (A) .. controls (0.05,0.3) and (0.1,0.5) .. (B)
                  .. controls (0.2,0.7) and (0.3,0.8) .. (C)
                  .. controls (0.4,0.88) and (0.5,0.88) .. (D)
                  .. controls (0.6,0.93) and (0.6,0.94) .. (E)
                  .. controls (0.9,0.99) and (0.95,1) .. (F);
        \draw[thick,->] (0,0) -- (1,0) node[right] {\footnotesize $x$};
        \draw[thick,->] (0,0) -- (0,1) node[above] {\footnotesize $P$};
    \end{scope}

    \begin{scope}[shift={(1.6cm, 2.2cm)}, scale=1]
  \coordinate (A) at (0,0);
        \coordinate (A) at (0,0);
        \coordinate (B) at (0.2,0.1);
        \coordinate (C) at (0.5,0.5);
        \coordinate (D) at (0.8,0.9);
        \coordinate (E) at (1,1);
        \draw (A) .. controls (0.05,0.05) and (0.15, 0.1) .. (B)
                  .. controls (0.4,0.3) and (0.5,0.45) .. (C)
                  .. controls (0.6,0.8) and (0.8,0.9) .. (D)
                  .. controls (0.95,0.9) and (0.98,1) .. (E);
        \draw[thick,->] (0,0) -- (1,0) node[right] {\footnotesize $x$};
        \draw[thick,->] (0,0) -- (0,1) node[above] {\footnotesize $P$};
    \end{scope}

    \begin{scope}[shift={(3.9cm, 2.2cm)}, scale=1]
    \node[draw=none] at (CDF1) {};
    \coordinate (A) at (0,0);
    \coordinate (B) at (0.2,0.1);
    \coordinate (C) at (0.4,0.2);
    \coordinate (D) at (0.6,0.5);
    \coordinate (E) at (0.8,0.8);
    \coordinate (F) at (1,1);

    \draw (A) .. controls (0.1,0.05) and (0.3,0.15) .. (B)
              .. controls (0.3,0.15) and (0.35,0.2) .. (C)
              .. controls (0.45,0.3) and (0.55,0.4) .. (D)
              .. controls (0.65,0.6) and (0.75,0.7) .. (E)
              .. controls (0.85,0.9) and (0.95,0.95) .. (F);
    \draw[thick,->] (0,0) -- (1,0) node[right] {\footnotesize $x$}; 
    \draw[thick,->] (0,0) -- (0,1) node[above] {\footnotesize $P$}; 
    \end{scope}

\end{tikzpicture}}
\caption{The \arcdf model: Autoregressively outputs parameters $\theta$ for monotonic rational-quadratic splines.}
\label{fig:arcdf}
\end{figure}

\subsection{Overview}
\noindent \textbf{The \textsc{NeuroCDF} Framework.}
\arcdf builds upon the \neucdf framework.
The central idea of \neucdf~\cite{wu2024practical} is to construct query-driven neural network (NN) models that predict the underlying cumulative distribution functions (CDFs) for cardinality estimation, rather than directly estimating query selectivities/cardinalities. Then, the selectivity of any rectangular query can be computed as a linear combination of the CDFs evaluated at its vertices~\cite{durrett2019probability}. Formally speaking, the CDF prediction model in \neucdf parameterizes the underlying CDFs: $ F(\mathbf{x})= P(X \leq \mathbf{x})$, where $\mathbf{x} = [{x}_1, {x}_2,..., {x}_d]$ is a real value variable $\mathbf{x}$, and $d$ is the dimension (\textit{i.e.,} the number of attributes of the table).

Compared to the common modeling paradigm that targets query cardinalities directly, \neucdf \emph{provably} provides robust cardinality estimates for out-of-distribution (OOD) queries. Specifically, it ensures the \emph{additivity constraint} or \emph{containment property} of cardinality functions: $\hat{c}(q_1) = \hat{c}(q_2) + \hat{c}(q_3)$ whenever $q_1 = q_2 \cup q_3$ and $q_2 \cap q_3 = \emptyset$. This property enhances \name's robustness to value distribution shifts for range predicates.

\smallskip
\noindent \textbf{Limitation.} Despite these advantages, \neucdf has a limitation. As noted in~\cite{wu2024practical}, \neucdf is not compatible with Q-error or SLE due to potential negative cardinality estimates. This issue arises because the multilayer perceptron (MLP) model used in \neucdf may fail to learn a monotonically increasing function ---  necessary property for CDFs. This restricts the practical applicability of \neucdf, as Q-error is a commonly used error metric in cardinality estimation due to their emphasis on highly selective queries~\cite{moerkotte2009preventing}.

\smallskip
\noindent \textbf{Our Solution.} We enhance the \neucdf framework by introducing an Autoregressive CDF prediction model. We refer to our solution as \arcdf: \neucdf with AutoRegressive CDF modeling, which is illustrated in Figure~\ref{fig:arcdf}. The key idea of \arcdf is to ensure that the CDF prediction model learns a monotonically increasing function \emph{by model design}, which alleviates negative cardinality estimates. This is achieved by: (1) employing a deep AR model to parameterize the CDFs, and (2) enforcing the monotonicity property along each attribute using \emph{monotonic piecewise splines}.

\subsection{Autoregressive CDF Modeling}

\arcdf decomposes the joint CDF $ F(\mathbf{x})= P(X \leq \mathbf{x})$ in an autoregressive manner:

\vspace{-3mm}
\begin{equation}
     F(\mathbf{x}) = \prod_{i=1}^{d}{P(X_i \leq x_i | X_{i-1} \leq x_{i-1}, \dots, X_1 \leq x_1 )},
\end{equation}
\vspace{-1mm}

where $P(X_i \leq x_i | X_{i-1} \leq x_{i-1}, \dots, X_1 \leq x_1 )$ is a \textit{conditional} CDF given the event: $X_{i-1} \leq x_{i-1}, \dots, X_1 \leq x_1$. For simplicity, we denote this as $P(X_i \leq x_i | \leq \mathbf{x_{1:i-1}})$ hereafter. This conditional CDF depends only on the values of $x_{1:i-1}$.  Unlike the conventional definition of conditional CDFs that condition on exact values (\textit{i.e.,} $X_{i-1} = x_{i-1}, \cdots , X_1 = x_1$), our definition conditions on \emph{inequalities}, which ensures the correctness of decomposition.

The motivation behind this autoregressive decomposition is twofold. First, it allows us to leverage recent advances in deep autoregressive models~\cite{germain2015made} for efficient and accurate (conditional) CDFs estimation. Second, by decomposing the joint CDF into a product of conditional CDFs, we can conveniently enforce the monotonicity constraint along each dimension.

\noindent \textbf{Modeling CDFs with Deep Autoregressive Models.} To parameterize the sequence of CDFs, we employ modern deep AR models such as MADE~\cite{germain2015made}. These models have proven to be efficient and powerful for cardinality estimation by parameterizing joint probability density functions (PDFs)~\cite{naru, yang2020neurocard}. To the best of our knowledge, we are the \emph{first} to leverage deep AR models to parameterize joint CDFs, enabling efficient and accurate CDF modeling.

Typically, deep AR models output a categorical distribution that represents the conditional probability distribution over the attribute domain at each dimension. Unfortunately, in the DACL problem, the domain values of each attribute are \emph{not} available.

\vspace{-0.5em}
\subsection{Parameterizing Conditional CDFs~\label{section.arcdf.splines}}

To address this challenge, \arcdf uses deep AR models to parameterize \emph{piecewise spline functions} along each attribute $x_i$, representing the conditional CDF, $P(X_i \leq x_i | \leq \mathbf{x_{1:i-1}})$. This approach offers two main advantages: (1) it allows for a flexible trade-off computational efficiency and model expressiveness without requiring knowledge of the attribute domain values; and (2) ensuring monotonicity in each conditional CDF becomes straightforward by leveraging the rich literature in monotonic piecewise spline functions~\cite{steffen1990simple,gregory1982piecewise}.

\noindent \textbf{Monotonic Rational-Quadratic Splines.}  
Common choices for monotonic piecewise spline functions include linear polynomial splines~\cite{rivlin1981introduction}, cubic splines~\cite{fritsch1980monotone}, and rational-quadratic splines~\cite{steffen1990simple}. Among these, rational-quadratic splines offer the most expressive modeling capability while maintaining computational efficiency, as demonstrated in applications such as density estimation and image generative modeling in recent ML research~\cite{durkan2019neural}. Therefore, \arcdf employs rational-quadratic splines to model the conditional CDFs for each attribute. Specifically, these splines are defined on one-dimensional data \( x \) and consist of \( K \) consecutive rational-quadratic functions, connected at \( K+1 \) monotonically increasing knots, denoted as \( \{(x^{k}, y^{k})\}_{k=0}^{K} \). The parameters for each segment are the outputs of the deep AR model at each dimension. Due to space constraints, we omit the exact mathematical expression of the rational-quadratic function; interested readers may refer to~\cite{gregory1982piecewise}.

\noindent \textbf{Validity of Learned Joint CDFs.}  \arcdf ensures the conditional CDFs are monotonically increasing because all knots are monotonically increasing and the derivatives $\delta$ are positive. However, in principle, this does not guarantee the validity of the learned joint CDFs. 
Monotonicity along each one-dimensional conditional CDF does not ensure that the cardinality estimate of the query over the multi-dimensional data is always valid  (\textit{i.e.,} non-negative). 

However, in our experiments, we find that \arcdf \emph{automatically} learns better CDFs, with only a small portion of ``outlier'' queries yielding negative estimates (\name switches to square error loss in this case). This improves on previous CDF model architectures (\textit{e.g.,} MLP).
We attribute this improvement to the autoregressive decomposition combined with the enforcement of monotonic conditional CDFs, allowing \arcdf to effectively learn valid joint CDFs from training queries. Although we cannot provide theoretical guarantees, our empirical results demonstrate that \arcdf achieves a significantly lower Q-error than other model choices in \neucdf. One could also add a penalty term to loss function to enforce the global monotonicity, but we do not explore this in the paper.

\vspace{-1mm}

\section{\mbox{Capturing Join Correlations}~\label{section.alv}}
In this section, we outline the practical challenges associated with using join key models $\mathcal{M}_{JK}$ in \name. We then present our proposed solution, which effectively mitigates these issues to ensure robust and efficient join correlation modeling.

\subsection{Challenges and the Idea}

\noindent \textbf{Challenges.} First, in real-world datasets, join keys often possess \textbf{large domain sizes}, with millions of unique values. This makes it highly impractical to build join key prediction models defined in \S~\ref{section.system.overview}. Second, as mentioned in the definition of the DACL problem, both the \textbf{join key domain and its size are unknown} due to no data access. This further complicates the task of constructing models that accurately capture join correlations.

\noindent \textbf{Key Idea.} \name leverages the notion of \emph{count sketches}~\cite{rusu2008sketches, alon1999tracking} to effectively approximate join sizes while accounting for join correlations. The fundamental concept behind count sketches involves creating \emph{low-dimensional} vector representations for data streams. For instance, consider two data streams \(S_1\) and \(S_2\) where each stream consists of tuples in the form (key, frequency), and the number of distinct key values is denoted as \(|\text{dom}(key)|\). By constructing compact vector representations \(\mathbf{v_{1}}\) and \(\mathbf{v_{2}}\) for \(S_1\) and \(S_2\) respectively (where $|\mathbf{v_{1}}| \ll |\text{dom}(key)|$), we can efficiently estimate the join size \(|S_1 \Join S_2|\) by computing the scalar dot product \(\mathbf{v_{1}} \boldsymbol{\cdot}\mathbf{v_{2}}\).

\subsection{Query-Driven Learned Count Sketches}
To estimate the join size of two query results, $q_1$ and $q_2$, with corresponding join key distributions $\mathbf{f}_1$ and $\mathbf{f}_2$, count sketches construct low-dimensional vectors $\mathbf{v}_1$ and $\mathbf{v}_2$ such that \(\Vert\mathbf{v_{1}} \boldsymbol{\cdot} \mathbf{v_{2}\Vert_1} \approx \Vert\mathbf{f_{1}} \cdot \mathbf{f_{2}}\Vert_1\). This approximation allows the join size to be estimated as $c(q_1) \cdot c(q_2) \cdot \Vert\mathbf{v}_1 \boldsymbol{\cdot} \mathbf{v}_2\Vert_1$. The approach naturally extends to joins involving more than two query results.

However, constructing count sketches typically requires scanning the underlying data, which is not permissible in the DACL problem. To overcome this limitation, we propose training machine learning models to generate \emph{learned} count sketches directly from input queries. These learned count sketch (LCS) models, denoted as $\mathcal{M}_{LCS}$, can seamlessly replace the join key models described in \S~\ref{section.compo_design}. The training process for $\mathcal{M}_{LCS}$ follows the same procedure as that for $\mathcal{M}_{JK}$ without accessing the underlying data. Therefore, we will use $\mathcal{M}_{LCS}$ to replace $\mathcal{M}_{JK}$ for each base relation, hereafter.

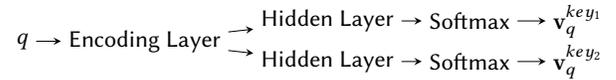
\begin{figure}[ht]
\scalebox{0.9}{
  \begin{tikzpicture}[>=stealth, every node/.style={font=\sffamily\large}]
    \node (q) at (0,0) {$q$};
    \node (encoding) at (1.8,0) {Encoding Layer};
    \node (hidden1) at (4.5,0.3) {Hidden Layer};
    \node (hidden2) at (4.5,-0.3) {Hidden Layer};
    \node (softmax1) at (6.6,0.3) {Softmax};
    \node (vq1) at (8.2,0.3) {$\mathbf{v}_{q}^{key_1}$};
    \node (softmax2) at (6.6,-0.3) {Softmax};
    \node (vq2) at (8.2,-0.3) {$\mathbf{v}_{q}^{key_2}$};

    \draw[->] (q) -- (encoding);
    \draw[->] (encoding) -- (hidden1);
    \draw[->] (hidden1) -- (softmax1);
    \draw[->] (encoding) -- (hidden2);
    \draw[->] (hidden2) -- (softmax2);
    \draw[->] (softmax1) -- (vq1);
    \draw[->] (softmax2) -- (vq2);
\end{tikzpicture}
  }
\caption{The Learned Count Sketch (LCS) model.} \label{fig:lcs}
\end{figure}

\noindent \textbf{Model Architecture.} For a single-table query $q$, the learned count sketch $\mathbf{v}_q$ is a normalized vector. 
The LCS model (Figure~\ref{fig:lcs}) encodes the query $q$ into an embedding/vector, processes it through a hidden layer, and outputs the learned count sketch $\mathbf{v}q$ (which is a $n_{lcs}$-dimensional normalized distribution) using a \texttt{softmax} activation function, where $n_{lcs}$ is a hyperparameter.
Existing query encoding techniques (e.g., MSCN~\cite{kipf2019learned} and flatten encoding~\cite{dutt2019selectivity}) can serve as the encoding layer. Again, \name chooses the MSCN encoding due to its flexibility in handling complex filters.  If the relation includes multiple join keys from different groups, the LCS models output a set of count sketches via \emph{different hidden layers}, all of which share a \emph{common encoding layer}.

\begin{exmp}\label{exmp.lcs}
Consider the LCS model in Figure~\ref{fig:lcs} where the table $T$ includes two different join keys, $key_1$ and $key_2$. Each count sketch is derived from the same query embedding but processed via different hidden layers for each key. Let's define the hidden layer size as $n_h$ (256 in our experiments). For each key, a hidden layer consists of a weight matrix $W \in \mathbb{R}^{n_h \times n_{lcs}}$ and a bias vector $b \in \mathbb{R}^{n_{lcs}}$. Thus, the model first produces an $n_h$-dimensional embedding $E_q$ from the encoding layer, and applies the transformation for the count sketch as $\mathbf{v}_q = \texttt{softmax}(E_q W + b)$. This transformation is executed separately for each key using distinct sets of $W$ and $b$.
\end{exmp}

\subsection{Comparison to Join Histograms~\label{section:lcs:compare}}
One may find that \name along with the LCS models share a similar high-level idea with join histograms~\cite{ioannidis1993optimal} (or FactorJoin~\cite{wu2023factorjoin} that extends the join histograms) --- \emph{learning per-table CardEst models and join correlation models separately}.

However, they are \emph{conceptually different} in three key ways.
\textbf{First}, join histograms are built from the data, whereas LCS models are learned exclusively from queries. \textbf{Second}, they rely on \emph{different assumptions}: join histograms assume that join keys within each bin are uniformly distributed, while LCS models assume that the dot products of join key distributions can be approximated using low-dimensional count sketches. \textbf{Third}, unlike the join histograms (or FactorJoin) framework which uses \emph{correlated} join histograms and per-table CardEst models for \emph{bin-specific} cardinality estimates, we employ a different approach due to production queries often lacking filters on join keys. Consequently, accurate bin-specific models are inaccessible. Instead, we utilize separate per-table CardEst models to estimate the \emph{overall} cardinality for joined tables without relying on binning. Additionally, LCS models are used to address the correlations between the results of per-table subqueries, \emph{disentangling} the estimation of per-table cardinalities and join correlations.

\section{Experimental Evaluation~\label{section.exps}}

In this section, we primarily answer: despite having no access to the data, \textbf{RQ1:} Can \name generalize to unseen join templates?  \textbf{RQ2:} Can \name maintain robustness to join template imbalance?  \textbf{RQ3:} Can \name achieve robustness to value distribution shifts in range predicates? Moreover, we also explore  \textbf{RQ4:} Can \name lead to better query end-to-end performance?

\vspace{-0.5em}
\begin{table}[ht!]
\caption{Statistics of data and workloads of DB instances}
\label{exp.table.stat}
\centering

\vspace{-0.4em}
\scalebox{0.9}{
\begin{tabular}{cccc}
  \hline
{DB Instance}  & \textbf{CEB-IMDb-full} &\textbf{DSB}  & {\color{black} \textbf{BD}}  \\
  \hline
  {\# Tables} & 21   & 25 & {\color{black}7}\\
{ Max \# tales in joins} & 16   & 5 & {\color{black}4} \\
\# All possible join templates & 3219   & 20 & {\color{black}28}\\
{{\color{black} Template Coverage Ratio (TCR)}} & \%10  & \%25  &  {\color{black}\%28}\\
{\color{black}Class Imbalance Ratio (CIR)} & {\color{black}108} & {\color{black}7.4} & {\color{black}17}\\
\hline
\end{tabular}}
\end{table}
\vspace{-1.2em}

\subsection{Experimental Setup\label{section:setup}}

\noindent \textbf{Datasets and Workloads.}
We conducted experiments using three DB instances: CEB-IMDb-full~\cite{negi2021flow}, DSB~\cite{ding2021dsb}, {\color{black} an internal business unit  (query logging system) at ByteDance (\textbf{BD}). This system stores information (\textit{e.g.,} count and running time) about each SQL template, and allows users to issue queries for monitoring and diagnosing slow queries.} Table~\ref{exp.table.stat} presents the statistics of data and workloads of these DB instances. The CEB-IMDb-full benchmark is derived from the JOB~\cite{leis2015good} benchmark, whose data distribution is highly correlated and skewed~\cite{leis2015good}. Moreover, CEB-IMDb-full provides hundreds of queries per handcrafted base template\footnote{The term "base template" refers to a predefined query template that generates various specific queries by substituting parameters within that template. 
} with real-world interpretations~\cite{negi2021flow} as opposed to five in the JOB benchmark. Moreover, CEB-IMDb-full contains various join types (\textit{e.g.,} star, chain, self-joins) and complex predicates (\textit{e.g.,} \texttt{LIKE}, \texttt{IN}). This variety makes it an ideal benchmark for our experiments. We primarily utilize it to evaluate  \textbf{RQ1} and {\color{black}\textbf{RQ2}} (due to its complexity), and \textbf{RQ4} (due to its large number of joins).
The DSB dataset is an extension of the TPC-DS benchmark~\cite{poess2002tpc} with more complex data distributions. Following~\cite{wu2024practical}, we populate the DSB dataset with a scale factor of 50 and use the workload generated by~\cite{wu2024practical} mainly for evaluating \textbf{RQ3} because its queries are range predicate-focused (though we also evaluate it for other RQs). {\color{black} For BD, we utilize real data and queries (over a day) directly collected from the internal business.}

\noindent \textbf{Baselines.} We are not aware of any deep learning-based query-driven approaches that operate without data access; non-deep learning approaches~\cite{hu2022selectivity, park2020quicksel} do not require data access, but they generally show poorer empirical performance~\cite{wu2024practical} and do not support joins. Therefore, we employ two representative deep-learning-based query-driven models, LW-NN~\cite{dutt2020efficiently} and MSCN~\cite{kipf2019learned}, as our baselines, modifying their query encodings to exclude data information.
We refer to the variants as $\text{LW-NN}_{\text{w/oData}}$ and $\text{MSCN}_{\text{w/oData}}$. 
We implemented LW-NN~\cite{dutt2019selectivity}. For MSCN, we utilized the available code from~\cite{mscncode}. Note that LW-NN's encoding is not flexible to handle $\texttt{LIKE}$ predicates; in such cases, we adopt MSCN's encoding while retaining LW-NN's approach to handling joins --- using a separate model for each join template. All query-driven models use a batch size of 128. For \name, the dimensions ($n_{lcs}$) of the LCS models are set to 2000, 100 and 3000 for CEB-IMDb-full, DSB and {\color{black}BD}, respectively.

Moreover, we include two unfair approaches that require data access for comparison: the histogram-based estimator in PostgreSQL (PG) to demonstrate the potential of query-driven models, and the original version of MSCN that incorporates sample bitmaps in its query encoding ($\text{MSCN}_{\text{w/Data}}$), to illustrate the superiority of \name. We do not include $\text{LW-NN}_{\text{w/Data}}$ as $\text{MSCN}_{\text{w/oData}}$ generally outperforms $\text{LW-NN}_{\text{w/oData}}$ in all experiments.


\noindent \textbf{Evaluation Metrics.} For prediction accuracy (\textbf{RQ1}, \textbf{RQ2}, and \textbf{RQ3}), we use Qerror. For query end-to-end performance (\textbf{RQ4}), we report the query running latency/time, {\color{black}which includes the query execution time and the inference time. The latter includes the total time the model takes to estimate the cardinalities for all subqueries.}

\noindent \textbf{Environment.} All model training was performed on a Nvidia Tesla V100 32GB GPU (although all models consume $<10\%$ of the GPU memory). All query latency experiments were performed on a Debian 9 Linux machine. The hardware included an Intel(R) Xeon(R) Platinum 8260 CPU @ 2.40GHz with 8 cores and a clock speed of 2.394 GHz, along with 16 GB of RAM. {\color{black}We utilize a modified version~\cite{han2021CEbenchmark} of PostgreSQL 13.1 for our query end-to-end performance evaluations. During query optimization, the PostgreSQL optimizer employs injected cardinality estimates for query planning.}

\begin{table*}[ht]
\caption{Accuracy (Qerror) on {\textcolor{darkblue}{seen}}/\uline{\textcolor{maroon}{unseen}} join templates, where \uline{\textcolor{maroon}{unseen}} join templates account for the majority of all possible join templates. The first two methods require data access, which is infeasible in the real-world setting of this paper.}\label{tab:unseen_accuracy}
\centering
\scalebox{0.87}{
\begin{tabular}{@{}lc|ccc|ccc|ccc@{}}
\toprule
\multirow{3}{*}{\textbf{Method}} &  \multirow{3}{*}{
  \begin{minipage}{1.4cm} 
  \centering 
    \textbf{Information} \\ 
     \centering 
   \textbf{Needed}
  \end{minipage}%
} & \multicolumn{3}{c}{\textbf{DSB}} & \multicolumn{3}{|c}{\textbf{CEB-IMDb-full}} & \multicolumn{3}{|c}{\textbf{{\color{black}BD (real workload)}}} \\ \cmidrule(l){3-11} 
                        & & Median     &  \%95  &  Mean   & Median     & \%95   &  Mean   & {\color{black}Median}     & {\color{black}\%95}   &  {\color{black}Mean}    \\ \midrule
\textbf{Postgres} &    Data         &  \textcolor{darkblue}{1.5}/\uline{\textcolor{maroon}{1.3}}  &  \textcolor{darkblue}{9.6}/\uline{\textcolor{maroon}{8.1}}   &     \textcolor{darkblue}{3.6}/\uline{\textcolor{maroon}{3.2}} 
 &  \textcolor{darkblue}{1781}/\uline{\textcolor{maroon}{6.1}} &  \textcolor{darkblue}{$1 \times  10^6$}/\uline{\textcolor{maroon}{702}} & \textcolor{darkblue}{$4 \times  10^5$}/\uline{\textcolor{maroon}{717}}  
 &  \textcolor{darkblue}{276}/\uline{\textcolor{maroon}{3.1}} 
 &  \textcolor{darkblue}{$1 \times  10^6$}/\uline{\textcolor{maroon}{184}} 
 & \textcolor{darkblue}{$9 \times  10^6$}/\uline{\textcolor{maroon}{$4 \times  10^4$}} 
        \\ \midrule
        
$\text{\textbf{MSCN}}_{\text{w/Data}}$ &  Queries + Data  & \textcolor{darkblue}{1.1}/\uline{\textcolor{maroon}{5.2}} & \textcolor{darkblue}{2.3}/\uline{\textcolor{maroon}{5533}} &  \textcolor{darkblue}{2.0}/\uline{\textcolor{maroon}{5915}} 
& \textcolor{darkblue}{1.3}/\uline{\textcolor{maroon}{1.8}}   &\textcolor{darkblue}{3.4}/\uline{\textcolor{maroon}{328}}  &  \textcolor{darkblue}{3.5}/\uline{\textcolor{maroon}{2188}} 
& \textcolor{darkblue}{1.3}/\uline{\textcolor{maroon}{32}}  
& \textcolor{darkblue}{3.2}/\uline{\textcolor{maroon}{1251}}  
& \textcolor{darkblue}{2.0}/\uline{\textcolor{maroon}{2048}} 
\\  \midrule  
$\text{\textbf{LW-NN}}_{\text{w/oData}}$  & Queries  &  \textcolor{darkblue}{2.3}/\uline{\textcolor{maroon}{n/a}} & \textcolor{darkblue}{37}/\uline{\textcolor{maroon}{n/a}} & \textcolor{darkblue}{7.2}/\uline{\textcolor{maroon}{n/a}} 
& \textcolor{darkblue}{3.1}/\uline{\textcolor{maroon}{n/a}} 
& \textcolor{darkblue}{42}/\uline{\textcolor{maroon}{n/a}} & \textcolor{darkblue}{17}/\uline{\textcolor{maroon}{n/a}}
& \textcolor{darkblue}{1.8}/\uline{\textcolor{maroon}{n/a}}  
& \textcolor{darkblue}{4.8}/\uline{\textcolor{maroon}{n/a}}  
& \textcolor{darkblue}{3.5}/\uline{\textcolor{maroon}{n/a}}  
      \\ \midrule
$\text{\textbf{MSCN}}_{\text{w/oData}}$      &     Queries      & \textcolor{darkblue}{1.5}/\uline{\textcolor{maroon}{1.9}} & \textcolor{darkblue}{5.3}/\uline{\textcolor{maroon}{$2 \times 10^6$}} &  \textcolor{darkblue}{2.1}/\uline{\textcolor{maroon}{$8 \times 10^4$}} & \textcolor{darkblue}{1.9}/\uline{\textcolor{maroon}{2.5}} & \textcolor{darkblue}{7.1}/\uline{\textcolor{maroon}{663}}  & \textcolor{darkblue}{3.7}/\uline{\textcolor{maroon}{2671}}
&\textcolor{darkblue}{1.6}/\uline{\textcolor{maroon}{75}} & \textcolor{darkblue}{4.0}/\uline{\textcolor{maroon}{1865}} & \textcolor{darkblue}{2.7}/\uline{\textcolor{maroon}{2904}}
        \\ \midrule
\textbf{\name}         &    Queries   & \textcolor{darkblue}{1.8}/\uline{\textcolor{maroon}{1.8}} & \textcolor{darkblue}{8.7}/\uline{\textcolor{maroon}{15}} & \textcolor{darkblue}{6.7}/\uline{\textcolor{maroon}{9.5}} 
& \textcolor{darkblue}{2.0}/\uline{\textcolor{maroon}{1.7}} & \textcolor{darkblue}{9.4}/\uline{\textcolor{maroon}{23}} & \textcolor{darkblue}{5.7}/\uline{\textcolor{maroon}{41}} & 
\textcolor{darkblue}{1.5}/\uline{\textcolor{maroon}{5.5}} & 
\textcolor{darkblue}{4.2}/\uline{\textcolor{maroon}{149}} & 
\textcolor{darkblue}{3.2}/\uline{\textcolor{maroon}{320}} \\
\bottomrule
\end{tabular}}
\end{table*}

\vspace{-1em}
\subsection{Generalization to Unseen Join Templates~\label{section:exp:unseen}}
This subsection evaluates \textbf{RQ1}. To construct the training workloads, we first retain queries from base templates, as these typically include the largest join templates observed in real-world workloads. From all the subqueries of these base templates, we randomly sample $10\% - 20\%$ of the join templates, and incorporate their queries into the training workload. {\color{black} For BD, we directly used real queries collected as the training workload}. The statistics of data and workloads {\color{black} (including TCR and CIR)} are shown in Table~\ref{exp.table.stat}.

Table~\ref{tab:unseen_accuracy} presents the accuracy of compared approaches for seen and unseen join templates, respectively. MSCN consistently outperforms LW-NN, demonstrating that MSCN's join handling approach provides a degree of generalizability superior to that of LW-NN. However, despite slight improvements from sample bitmaps {\color{black}(which are independently sampled per table and lack information on join correlations)}, \mscnwd's performance \emph{significantly} deteriorates when moving from seen to unseen join templates. Notably, it even underperforms PG in terms of mean Q-error on unseen join templates within the CEB-IMDB-full benchmark. {\color{black} Note that PG is data-driven and does not require training queries. This is why its Qerror is worse for `seen' versus `unseen' queries, as the seen set contains the largest join templates, which typically pose challenges for PG.}

\emph{Importantly, we observe \name can achieve very consistent and robust accuracy on both seen and unseen join templates, across the three datasets.} Moreover, it even surpasses $\text{MSCN}_{\text{w/Data}}$ and PG that are built over the data. This demonstrates \name's superior generalization performance, attributed to its \emph{compositional design.}

\subsubsection{\textbf{Ablation Study}}
\begin{figure}
  \centering
  \begin{subfigure}[b]{0.235\textwidth}
      \includegraphics[width=\textwidth]{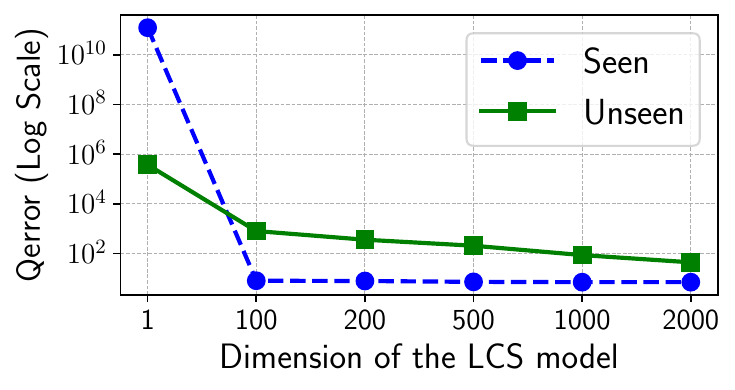}
         \captionsetup{skip=2pt}
      \caption{CEB-IMDb-full}
  \end{subfigure}
  \hfill
  \begin{subfigure}[b]{0.235\textwidth}
      \includegraphics[width=\textwidth]{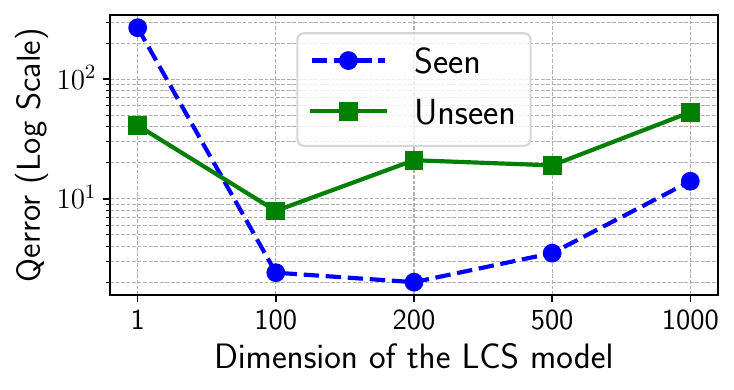}
         \captionsetup{skip=2pt}
      \caption{DSB}
  \end{subfigure}
  \caption{Mean Qerror with varying dimensions of the learned count sketch model across two datasets.}
  \label{fig:ablation_lcs}
\end{figure}

We also perform an ablation study to validate the effectiveness of the proposed learned count sketch model.
Figure~\ref{fig:ablation_lcs} presents the estimation accuracy of \name for seen and unseen join templates across different dimensions ($n_{lcs}$) of LCS models. The significant improvement from $n_{lcs}=1$ (representing independent join assumptions) to $n_{lcs}=100$ confirms that \emph{LCS models effectively capture join correlations among relations}.

\begin{figure}
  \centering
  \begin{subfigure}{0.23\textwidth}
    \includegraphics[width=\textwidth]{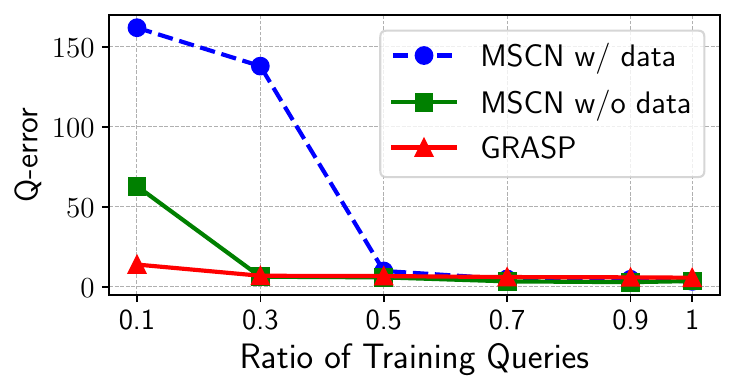}
    \captionsetup{skip=2pt}
    \caption{\color{black}{Seen Join Templates}}
    \label{subfig:complexity:in}
\end{subfigure}
  \begin{subfigure}{0.23\textwidth}
    \includegraphics[width=\textwidth]{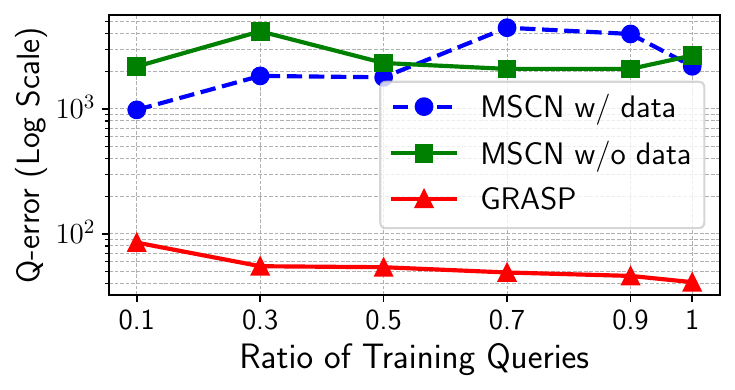}
    \captionsetup{skip=2pt}
    \caption{{\color{black}Unseen Join Templates}}
    \label{subfig:complexity:ood}
  \end{subfigure}
  \caption{{\color{black}Mean Q-error with varying ratios of training queries on CEB-IMDb-full.}}
  \label{fig:complexity}
\end{figure}

\begin{figure}
  \centering
  \begin{subfigure}{0.23\textwidth}
    \includegraphics[width=\textwidth]{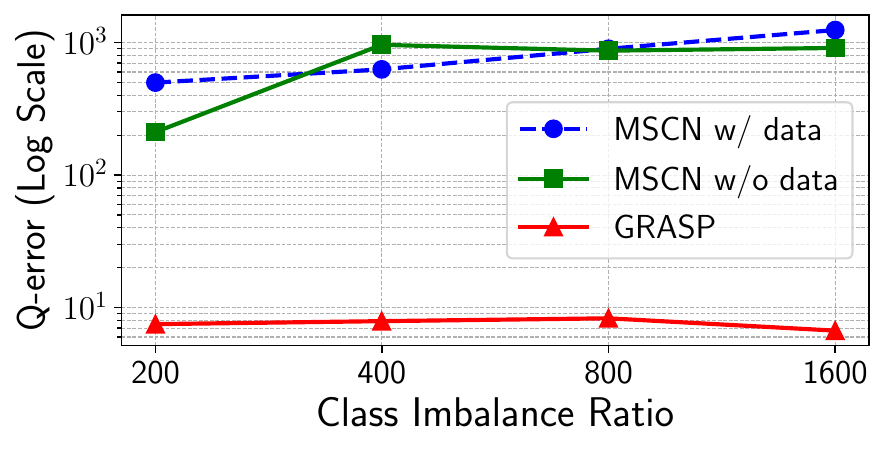}
       \captionsetup{skip=2pt}
    \caption{CEB-IMDb-2}
    \label{subfig:imb:ceb1}
  \end{subfigure}
  \hfill
  \begin{subfigure}{0.23\textwidth}
    \includegraphics[width=\textwidth]{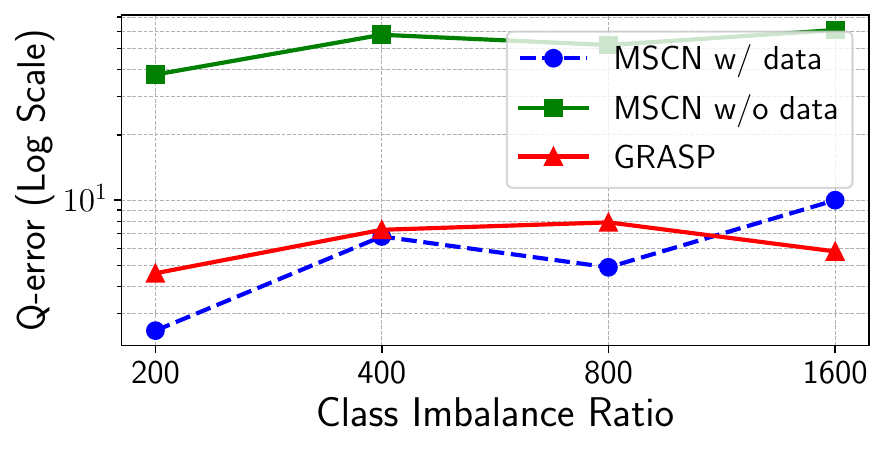}
       \captionsetup{skip=2pt}
    \caption{DSB}
    \label{subfig:imb:dsb}
\end{subfigure}
  \caption{Accuracy with varying class imbalance ratios (CIR).}
  \label{fig:imb_accuracy}
\end{figure}

\subsubsection{\textbf{Sample Complexity}\label{section:complexity}} 

We then evaluate the sample complexity of \name and two MSCN variants. We replicate the experiments in \S~\ref{section:exp:unseen}, varying the training query ratios. Figure~\ref{fig:complexity} shows the mean Q-errors for seen and unseen join templates at different ratios.
Notably, \name achieves good accuracy with only 10\% (25,701) of the original training queries for both seen and unseen join templates, suggesting its \emph{moderate} sample complexity. Additionally, \name consistently outperforms both MSCN variants across all training query ratios for unseen queries. 

\subsection{Robustness to Join Template Imbalance~\label{section:exp:imb}}

Now, we assess the robustness of the compared methods to join template imbalances (\textbf{RQ2}). We employ CIR~\cite{thabtah2020data} (as defined in Definition~\ref{def:cir}) to quantify the imbalance scale. To create imbalanced workloads for a specified CIR, we first sort the join templates in descending order based on their query counts. Starting with the largest count ($n_l$), we iteratively reduce each subsequent template's query count by a decay factor (e.g., 1.5) until the condition $n_l/n_s \geq \text{CIR}$ is satisfied, where $n_s$ is the current query count. All remaining join templates are then assigned a query count of $n_s$. We evaluate their performance on test workloads that contain an \emph{equal} number of queries for each join template to assess robustness to join template imbalance.
For CEB-IMDb-full, we create imbalanced workloads by combining base templates that share the same prefix ID (e.g., 3a, 3b) with their respective subqueries. This enables a more fine-grained evaluation of robustness due to the extensive variety of join templates in CEB-IMDb-full. For DSB, we directly construct imbalanced workloads from all available join templates, as it contains a considerably smaller number of join templates.

Figure~\ref{fig:imb_accuracy} presents the accuracy of query-driven approaches, excluding LW-NN since it is consistently outperformed by MSCN. CEB-IMDb-2 is constructed using base templates with prefix IDs 2 from CEB-IMDb-full. As we can see, on DSB, \mscnwd improves upon \mscnnod and is comparable to \name. However, when evaluating on CEB-IMDb-2, which features more complex join graphs, \name \emph{significantly} outperforms both MSCN variants. This demonstrates that \emph{\name is robust to join template imbalance.}

\subsection{Robustness to Value Distribution Shifts\label{section:exp:value}}

\begin{figure}
  \centering
  \begin{subfigure}[b]{0.23\textwidth}
      \includegraphics[width=\textwidth]{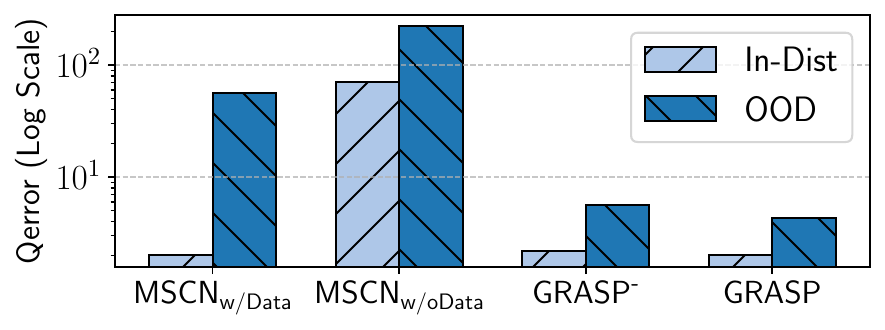}
         \captionsetup{skip=2pt}
      \caption{{\color{black}Card. Shift - Mean}}
  \end{subfigure}
  \begin{subfigure}[b]{0.23\textwidth}
      \includegraphics[width=\textwidth]{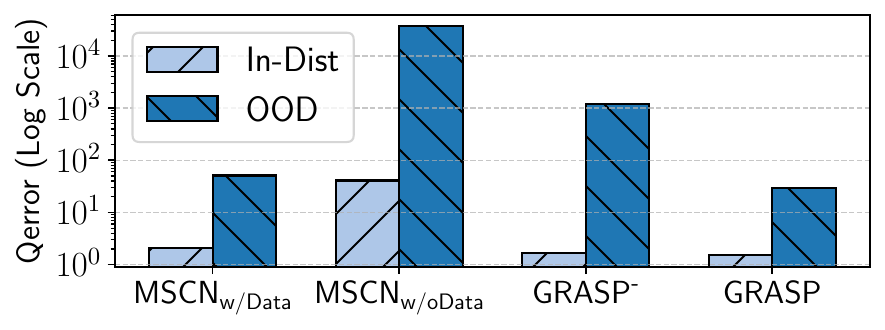}
         \captionsetup{skip=2pt}
      \caption{{\color{black}Gran. Shift - Mean}}
  \end{subfigure}
  \vspace{-0.5em}
  \caption{{\color{black}Accuracy under value distribution shifts for DSB.}}
  \label{fig:ood_accuracy}
\end{figure}

This subsection addresses \textbf{RQ3} using the DSB dataset. Following~\cite{wu2024practical}, we introduce granularity (Gran.) shifts, where the granularity ($g$) of range predicates changes {\color{black}(from $g \neq 0.5$ to $g =0.5$)}. Additionally, we evaluate cardinality (Card.) shifts, referring to changes in query cardinalities (e.g., from small to large values).

Figure~\ref{fig:ood_accuracy} presents the  mean Q-errors of query-driven approaches under both shift types. {\color{black} Here, $\text{GRASP}^-$ represents \name using the original \neucdf as the CDF prediction model (\textit{i.e.,} \name without \arcdf). We use $\text{GRASP}^-$ to evaluate the accuracy of \neucdf and to determine if \arcdf improves on \neucdf.

As shown in the figure, \mscnwd improves MSCN's robustness to value distribution shifts, especially for granularity shifts. Furthermore, $\text{GRASP}^-$ consistently outperforms \mscnnod, \emph{indicating that \neucdf is more robust to value distribution shifts compared to \mscnnod.} However, for granularity shifts, $\text{GRASP}^-$ is less effective than \mscnwd due to latter's additional data information. The gap can be bridged by \arcdf (used in $\text{GRASP}$), which demonstrates \arcdf's advantages over \neucdf. 

\subsection{Impact on End-to-End Performance\label{section:end2end}}

This subsection evaluates the benefits of \name in improving query latency performance (\textbf{RQ4}). We focus on the more complex CEB-IMDb-full (with up to 16-way joins) and {\color{black} real queries in BD}. 
For the test workloads, we \emph{randomly} sample four workloads from all base templates (\textit{e.g.,} those with larger joins), consisting of 10 queries each for CEB-IMDB-full and 20 queries each for BD. We also include True Cardinalities (True-Card) as Oracle. {\color{black} Note that compared approaches use the same
runtime, differing only in the cardinality estimates.}

\begin{figure}
  \centering
  \begin{subfigure}[b]{0.23\textwidth}
      \includegraphics[width=\textwidth]{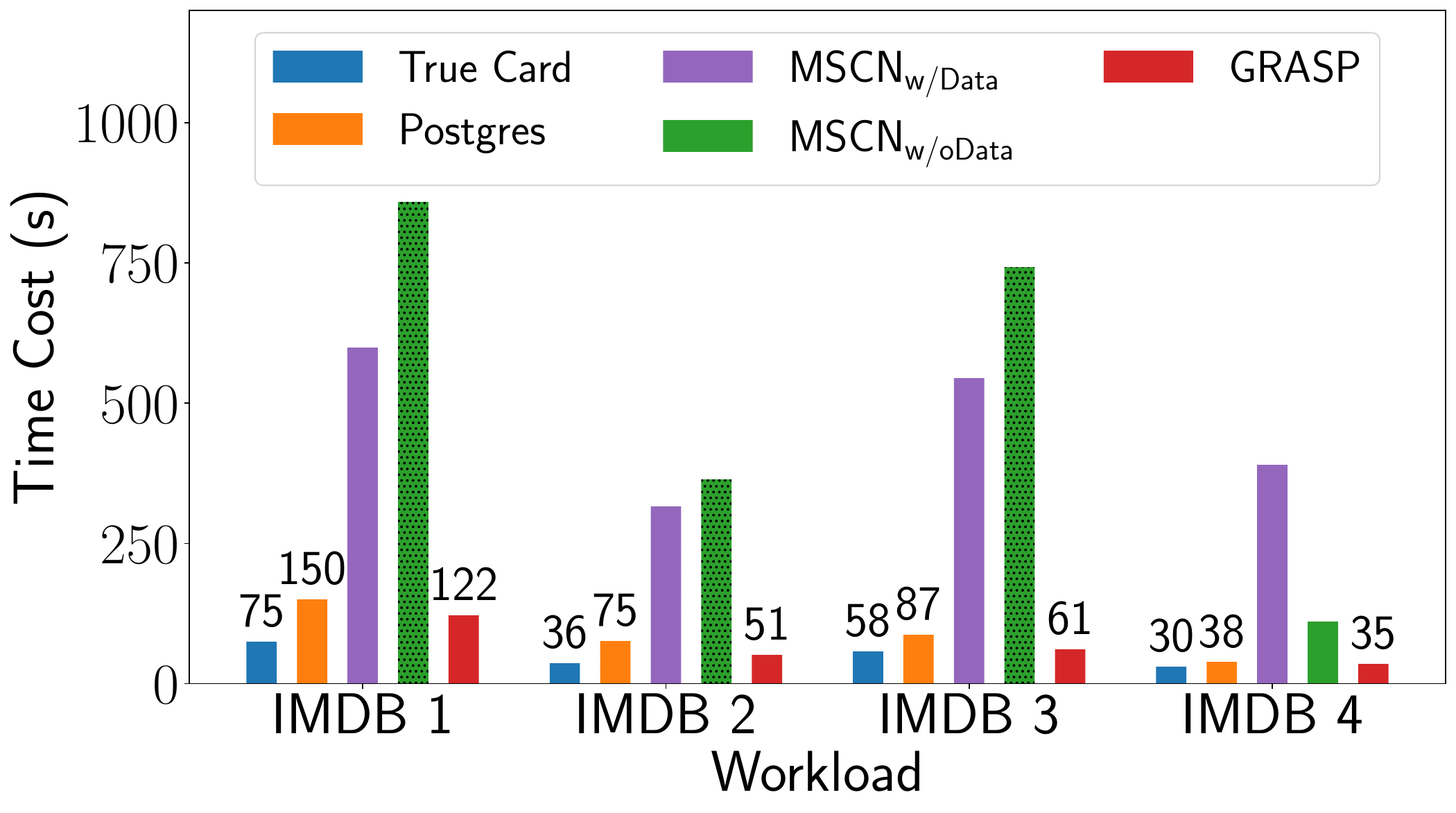}
         \captionsetup{skip=2pt}
      \caption{CEB-IMDb-full}  \label{fig:imdb_qo_comparison}
  \end{subfigure}
  \hfill
  \begin{subfigure}[b]{0.23\textwidth}
      \includegraphics[width=\textwidth]{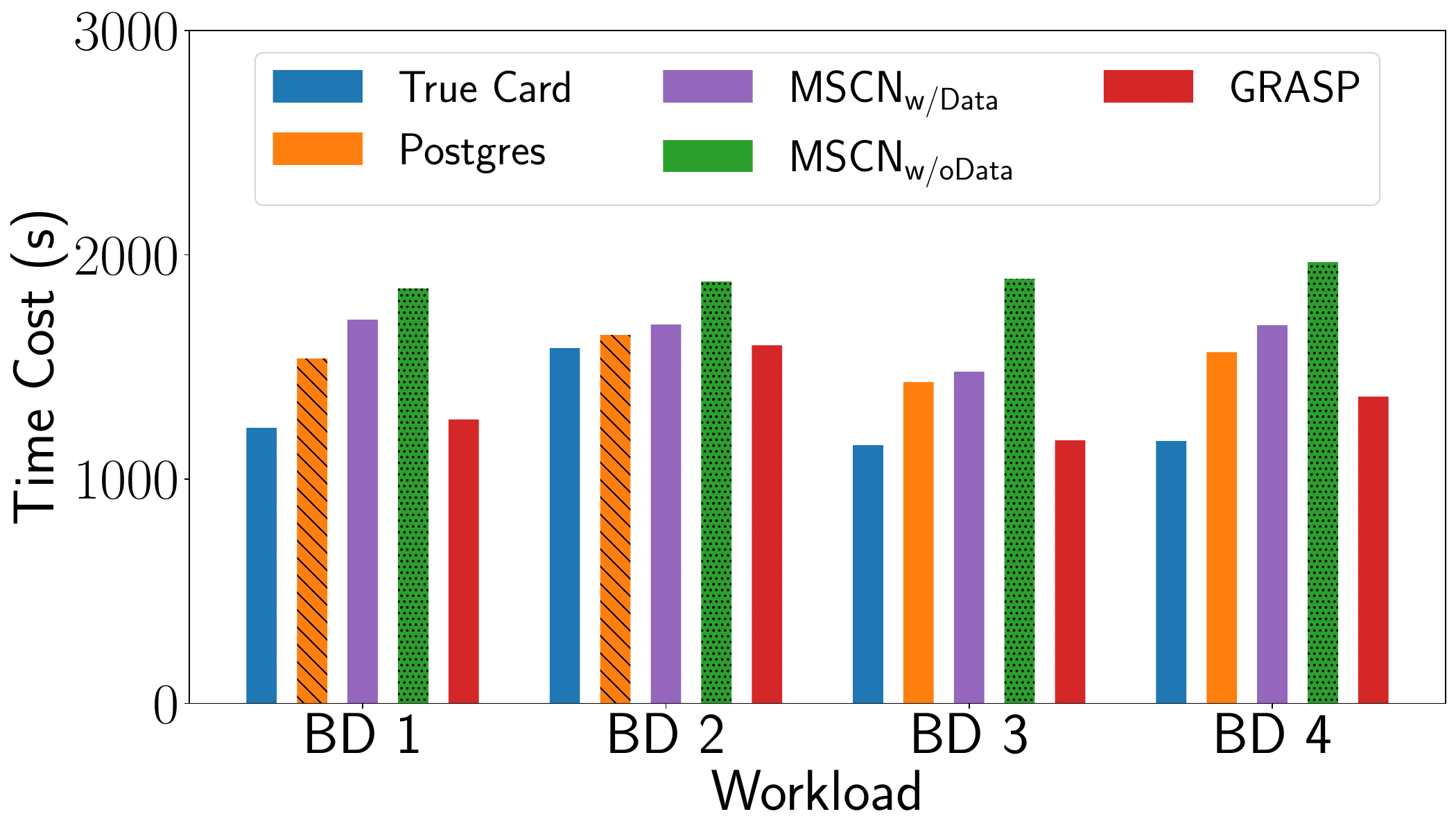}
         \captionsetup{skip=2pt}
\caption{{\color{black}BD}}\label{fig:bd_qo_comparison}
  \end{subfigure}
  \vspace{-0.5em}
  \caption{Overall time cost (running time).}
  \label{fig:imdb_dsb_qo_comparison}
\end{figure}

\begin{figure*}
    \centering
    \begin{subfigure}[b]{0.48\textwidth}
        \includegraphics[width=\textwidth]{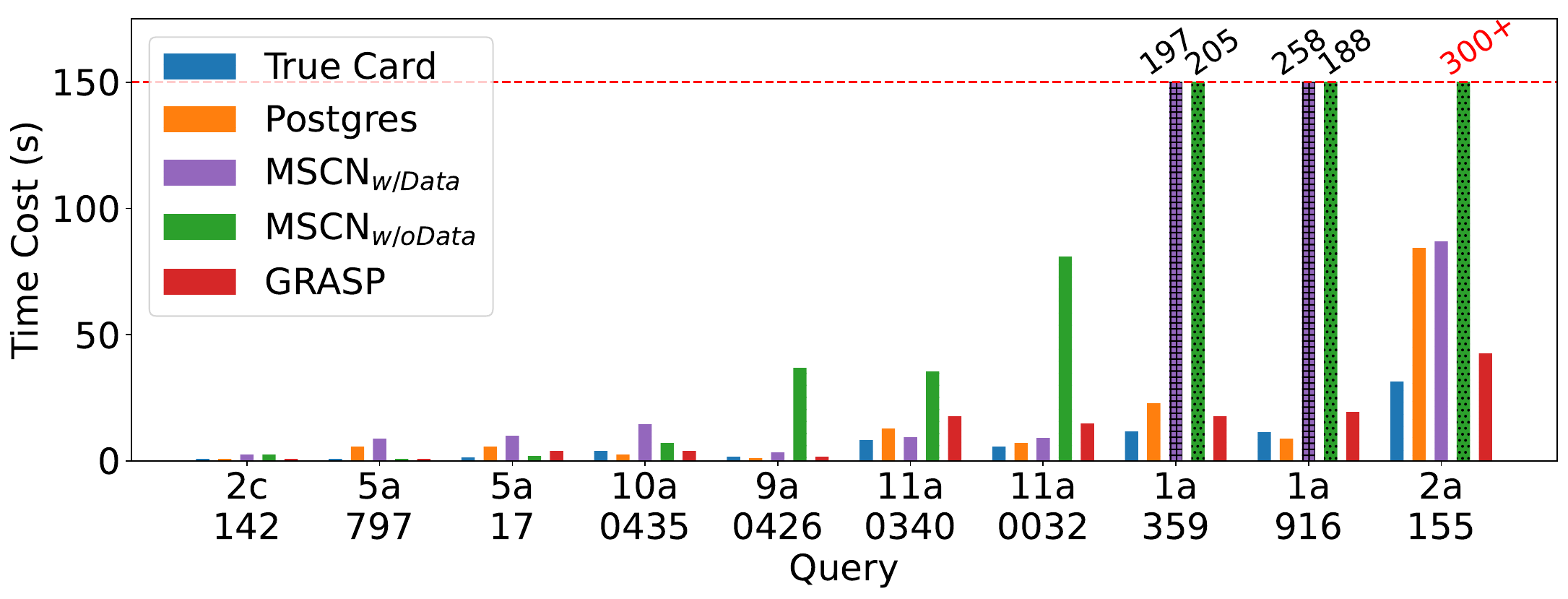}
        \caption{CEB-IMDb-full - Workload 1}
    \end{subfigure}
    \hfill
    \begin{subfigure}[b]{0.48\textwidth}
        \includegraphics[width=\textwidth]{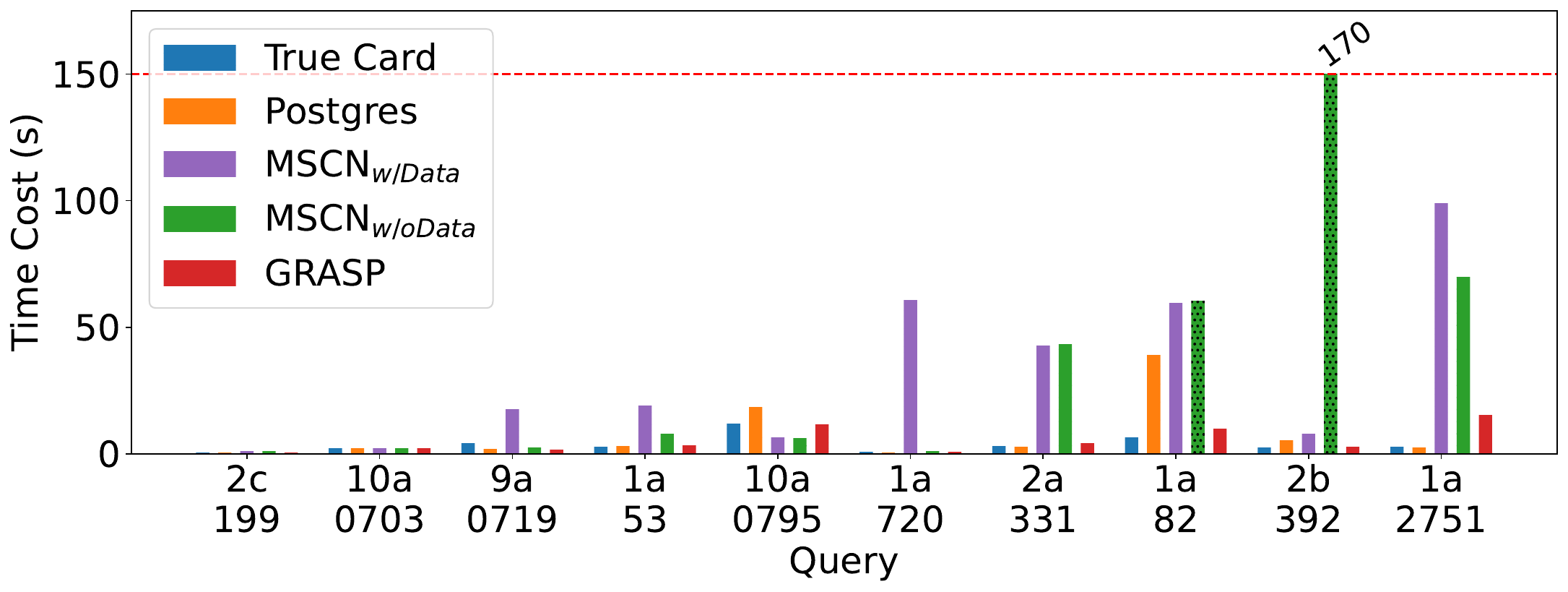}
        \caption{CEB-IMDb-full -Workload 2}
    \end{subfigure}
    
    \begin{subfigure}[b]{0.48\textwidth}
        \includegraphics[width=\textwidth]{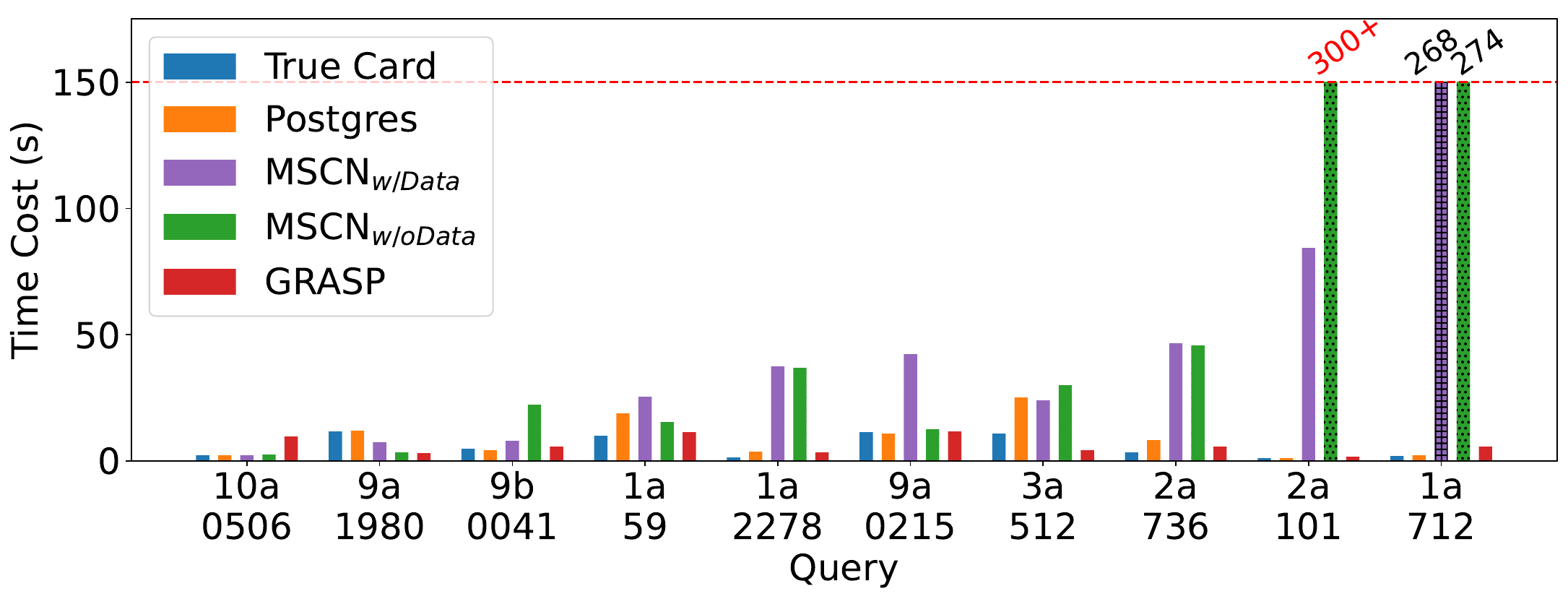}
        \caption{CEB-IMDb-full - Workload 3}
    \end{subfigure}
    \hfill
    \begin{subfigure}[b]{0.48\textwidth}
        \includegraphics[width=\textwidth]{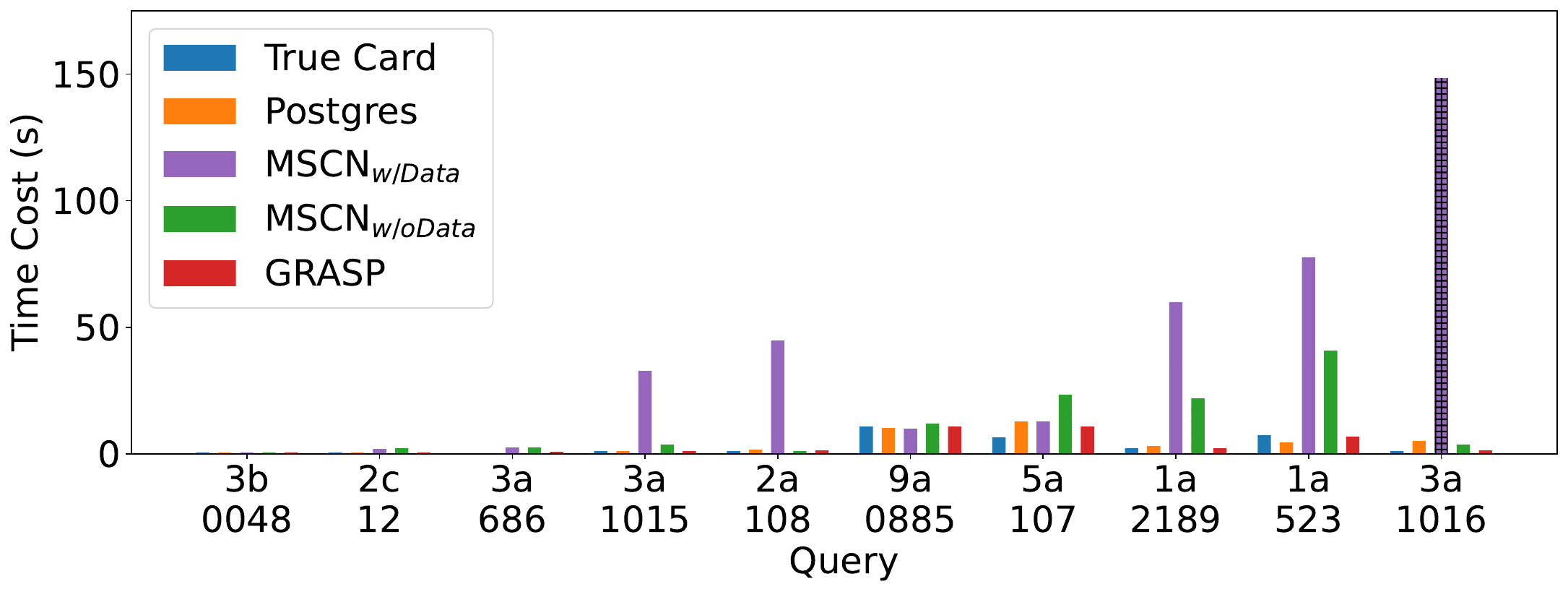}
        \caption{CEB-IMDb-full - Workload 4}
    \end{subfigure}
    \caption{Per-query running time performance on CEB-IMDb-full}
    \label{fig:imdb_qo}
\end{figure*}

{\color{black} Figure~\ref{fig:imdb_dsb_qo_comparison} 
presents the overall running time for both datasets. We report the ratios of the total running time (summing all four workloads) of the evaluated models relative to True-Card, which serves as the baseline and is set at 100\%.}  On CEB-IMDb-full, \name incurs a running time ratio of 135\%, significantly outperforming $\text{MSCN}_{\text{w/oData}}$ (1044\%) and $\text{MSCN}_{\text{w/Data}}$ (930\%), while achieving performance better than PG (176\%). It is worth noting that both PG and $\text{MSCN}_{\text{w/Data}}$ have access to full data statistics. These results demonstrate \emph{the strong performance of \name in enhancing query optimization.}
{\color{black}On BD, GRASP achieves running time performance comparable to True-Card, while significantly outperforming  $\text{MSCN}_{\text{w/Data}}$ and $\text{MSCN}_{\text{w/oData}}$ (126\%).}
This suggests that \emph{query-driven models may perform worse than simple baselines like PG when they fail to accurately estimate the majority of subqueries.}

Figure~\ref{fig:imdb_qo} shows the per-query runtime for the four workloads on CEB-IMDb-full, with a timeout of 300 seconds. We observe that both $\text{MSCN}_{\text{w/Data}}$ and $\text{MSCN}_{\text{w/oData}}$ time out on some queries. Additionally, $\text{MSCN}_{\text{w/oData}}$ exhibits significant degradation on a large number of queries.
 In contrast, \name achieves running times close to True-Card for the vast majority of queries without noticeable query regression. This demonstrates that \name effectively generalizes to various join templates, \emph{despite being trained on only 10\% of all possible join templates.}




\subsection{Efficiency\label{section:exp:efficiency}}

\noindent \textbf{Training Time.} Training \name takes 198s, 4.9s, and {\color{black} 51s} per epoch on the IMDB-CEB-full, DSB, and {\color{black} BD  datasets, respectively.  Moreover, \name converges within 50 epochs for all datasets.

\noindent \textbf{Query Inference.} We focus on the query inference of the largest join (16-way join) on CEB-IMDb-full. For query optimization of such a 16-way join query, \name completes the estimation of all subqueries (including itself) in less than 0.5s using batch inference on a GPU. {\color{black} On a CPU, \name achieves 1.6s inference time, which is not costly compared to the query running time. Specifically, although a 16-way join query involves estimating 6092 subqueries, only 32 calls to \name's primitive models (CardEst and LCS models) are needed --- one call to each model for each of the 16 tables. This step only takes 0.15s. After obtaining these model outputs, the cardinality of each subquery is estimated using these outputs (Algorithm~\ref{alg.cardset}). Here, the computations for the 6092 subqueries can be efficiently handled through progressive query inference. For instance, the cardinality estimate for $\texttt{t} \Join \texttt{mi}$ is cached and subsequently reused in the estimation of $\texttt{t} \Join \texttt{mi} \Join \texttt{ci}$.
}





\section{Related Work\label{sec:related}}

CardEst dates back to the early days of query optimization~\cite{selinger1979access,lynch1988selectivity}. Early methods relied on data statistics, such as histograms~\cite{deshpande2001independence,gunopulos2000approximating, gunopulos2005selectivity, muralikrishna1988equi}, assuming uniformity within buckets and independence across columns. However, it often results in significant estimation errors~\cite{ioannidis1991propagation,leis2015good}. Such techniques were refined to use queries themselves to compute histograms~\cite{aboulnaga1999self, bruno2001stholes, lim2003sash},  query expression statistics~\cite{bruno2002exploiting} and adjustments to correlated predicates~\cite{markl2003leo}

Recently, CardEst has been approached as an ML problem. ML-based CardEst methods are categorized into \emph{data-driven} and \emph{query-driven} models (with a few exceptions of hybrid approaches~\cite{wu2021unified, zhang2024duet, li2023alece}).
 Data-driven techniques~\cite{yang2020neurocard, naru, kim2024asm, heddes2024convolution, deeds2023safebound, deepdb} build models of the data distribution by scanning the underlying data.
 
Query-driven techniques fall into two main categories. The first constructs a \emph{data model} based on observed queries and their cardinalities~\cite{aboulnaga1999self, hu2022selectivity, park2020quicksel}. Although these methods generally do not require data access, they often perform worse than regression-based models due to limited capacity. The second category, initiated by~\cite{anagnostopoulos2017query,anagnostopoulos2015learning,anagnostopoulos2015learningidcm} (without join support), learns a \emph{regression model} from query features to cardinality. Recent deep learning-based approaches~\cite{kipf2019learned, dutt2020efficiently} in this category support joins and demonstrate impressive empirical performance. However, they require data access for improved generalizability and assume training workloads with complete and balanced join templates.

\section{Conclusions}
We introduced a \textit{new problem setting} for CardEst: \emph{data-agnostic cardinality learning from imperfect workloads}, grounded in real-world scenarios and a detailed analysis of production workloads. To solve this challenging problem, we developed \name, a truly \textit{data-agnostic} CardEst system that handles incomplete and imbalanced join templates through its \textit{compositional design}. Additionally, we proposed a query-driven CardEst model to address value distribution shifts for range predicates, and a novel \emph{learned count sketch} model that efficiently captures join correlations across base relations. We empirically validated the generalizability and robustness of \name using three database instances.

\begin{acks}
Components of this work were funded by NSF grant DBI-2400135.
\end{acks}

\bibliographystyle{acm}
\bibliography{main}
\end{document}